\documentclass[doublecol]{epl2} 

\usepackage{pdfpages} 
\usepackage{graphicx}
\usepackage{amssymb}
\usepackage{amsmath}
\usepackage[utf8]{inputenc}
\usepackage{cancel}
\usepackage{psfrag}
\usepackage{epsfig}
\usepackage{bbm}
\usepackage{bm}
\usepackage{dsfont}
\usepackage{soul}
\usepackage{colonequals}
\usepackage{float}
\usepackage{enumitem}
\usepackage{appendix}
\usepackage{lmodern}
\usepackage[T1]{fontenc}
\usepackage{physics}                                                
\usepackage{hyperref}
\usepackage[capitalize]{cleveref}
\usepackage{xcolor}								                 
\usepackage{xstring}                                                

\newcommand{\ccite}[1]{\IfSubStr{#1}{,}{Refs.~}{Ref.~}\cite{#1}}    
\newcommand{\cor}[1]{\mathcal{#1}}									
\newcommand{\T}[1]{\text{#1}}										
\newcommand{\dslash}[1]{\frac{\dd[d]{#1}}{(2\pi)^d}}                

\newcommand{\ie}{i.e.}
\newcommand{\n}{\nonumber}

\newcommand{\rev}[1]{#1}
\newcommand{\revv}[1]{#1}

\newcommand{\avg}[1]{\left\langle #1 \right \rangle}

\newcommand{\dbar}{{\mkern+3mu\raisebox{-0.6pt}{$\mathchar'26$}\mkern-11mu {\rm d}}}

\newcommand{\ti}{t_\mathrm{i}}
\newcommand{\tf}{t_\mathrm{f}}
\renewcommand{\a}{\mathfrak{a}}

\definecolor{DarkRed}{RGB}{153,0,0}
\definecolor{DarkGreen}{RGB}{0,153,0}
\definecolor{DarkBlue}{RGB}{0,0,153}

\newcommand{\out}[1]{ }

\newcommand{\Hp}{\mathcal{H}_\phi}

\newcommand{\Hint}{\mathcal{H}^\mathrm{int}}

\newcommand{\p}{\phi}
\newcommand{\la}{\left\langle}
\newcommand{\ra}{\right\rangle}
\newcommand{\dwint}{\dbar  W^\mathrm{int}}

\newcommand{\dqy}{\dbar Q_y}
\newcommand{\dqp}{\dbar Q_\p}
\newcommand{\dq}{\dbar Q}
\newcommand{\dw}{\dbar W}

\title{Stochastic thermodynamics of a probe in a fluctuating correlated field}

\author{Davide Venturelli\inst{1,2,6}\thanks{\email{davide.venturelli@sissa.it}} \and Sarah~A.~M.~Loos\inst{3,4,6}\thanks{\email{sl2127@cam.ac.uk}} \and Benjamin Walter\inst{5,1,6}\thanks{\email{b.walter@imperial.ac.uk}} \and \'Edgar Rold\'an\inst{4} \and Andrea Gambassi\inst{1}}
\shortauthor{Davide Venturelli \etal} 

\institute{                    
  \inst{1} SISSA --- International School for Advanced Studies and INFN, via Bonomea 265, 34136 Trieste, Italy\\
  \inst{2} Laboratoire de Physique Théorique de la Matière Condensée, CNRS/Sorbonne Université, 75005 Paris, France\\
  \inst{3} DAMTP, Centre for Mathematical Sciences, University of Cambridge, Wilberforce Rd, CB3 0WA Cambridge, United Kingdom\\
  \inst{4} ICTP --- International Centre for Theoretical Physics, Strada Costiera 11, 34151 Trieste, Italy\\
  \inst{5} Department of Mathematics, Imperial College London, 180 Queen's Gate, SW7 2AZ London, United Kingdom\\
  \inst{6} These authors contributed equally to this work.
}

\abstract{We develop a
framework for the stochastic thermodynamics of a probe coupled to a fluctuating medium with spatio-temporal correlations, described by a scalar field. For a Brownian particle dragged by a harmonic trap through a fluctuating Gaussian field, we show that near criticality 
\revv{(where the field displays long-range spatial correlations)} 
the spatially-resolved \revv{average} heat flux 
develops a 
dipolar structure, 
where heat is 
absorbed in front and dissipated behind the \revv{dragged} particle.
Moreover, a perturbative calculation reveals that the dissipated power displays three distinct dynamical regimes depending on the drag velocity.  
}

\begin{document}

\maketitle

Stochastic thermodynamics provides powerful tools to investigate the entropic and energetic properties of fluctuating systems 
coupled to media, even far from equilibrium~\cite{seifert2012stochastic,sekimoto2010stochastic,van2015ensemble,peliti2021stochastic,BO20171,roldan2022martingales}. A crucial
result in this context is the quantitative relation between the time-reversal asymmetry of the system's fluctuations 
and the heat dissipated to the environment,
associated 
with
entropy 
production~\cite{maes2003time,gaspard2004time,parrondo2009entropy,roldan2010estimating}.
Most of the previous works relied on the assumption
 that the environment is at all times in equilibrium  
 or 
 in a nonequilibrium steady state, and displays no dynamical spatio-temporal correlations. 
In various contexts, however, the dynamics crucially hinges on spatio-temporal correlations of the environment
--- this is the case, e.g., for
inclusions in lipid membranes~\cite{reister_lateral_2005,reister-gottfried_diffusing_2010,camley_contributions_2012,camley_fluctuating_2014}, microemulsions~\cite{Gompper_1994,Hennes_1996,Gonnella_1997,Gonnella_1998},
or defects in ferromagnetic systems~\cite{demery2010, demery2010-2, demery2011, demerypath, Dean_2011, demery2013}. 
 These
 correlations become long-ranged and particularly relevant when the environment is close to a critical point, as in the case, e.g., of
 colloidal particles in binary liquid mixtures~\cite{Hertlein_2008,Gambassi_2009,volpe2011microswimmers,Paladugu_2016,Ciliberto_2017,Magazzu_2019,martinez2023laserinduced}.
 Moreover, 
the simplified assumption of a structureless environment implies
that all the information about local energy and entropy flows occurring within the environment is 
not taken into account~\cite{Seifert_2016}. 
To overcome this paradigm, there is growing interest in extending concepts from stochastic thermodynamics towards systems with spatially-extended correlations~\cite{Leonard_2013,Niggemann_2020}, such as 
pattern-forming many-body systems~\cite{falasco2018information,suchanek2023entropy,pruessner2022field,suchanek2023irreversible}
or
critical media~\cite{campisi2016power,holubec2017work,herpich2020stochastic,caballero2020stealth}.
Furthermore, a
line of recent works discusses the irreversibility of active many-particle systems described by hydrodynamic field theories~\cite{li2021steady,nardini2017entropy,markovich2021thermodynamics,caballero2020stealth,suchanek2023irreversible}.
These theoretical advances are driven by state-of-the-art experiments involving optical trapping, ultrafast video-microscopy, or active particles~\cite{di2023variance,bechinger2016active,ro2022model,battle2016broken,Mestres_2014}.

\begin{figure}
    \centering
    \begin{flushleft}
    \vspace*{0.3cm}~~\\
    (a)\vspace*{-0.9cm}\\
    \end{flushleft}
    \includegraphics[width=0.7\columnwidth]{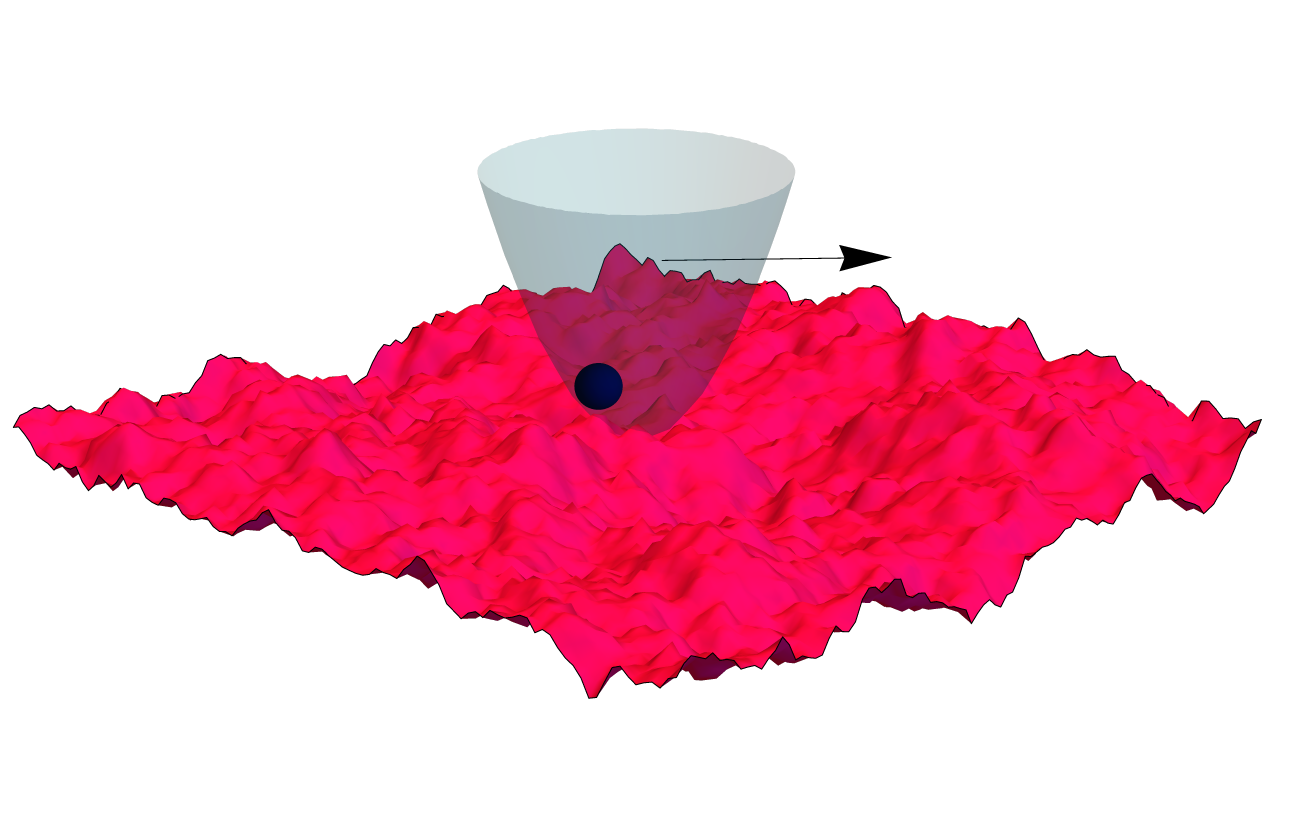}
    \put(-55,80){$\vb v$} 
    \begin{flushleft}
    (b)\vspace*{-1.1cm}\\
    \end{flushleft}
    \includegraphics[width=0.85\columnwidth]{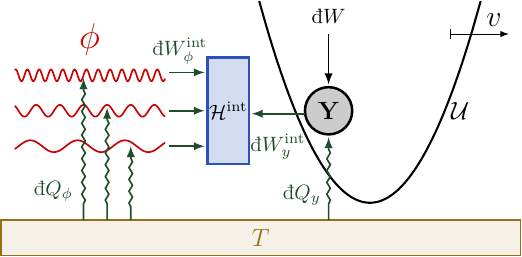}
    \caption{
    \textbf{(a)} Sketch of a probe particle dragged by a trap with 
    velocity $\vb v$ through a correlated field (red 
    surface). \textbf{(b)} \rev{Cartoon of the}
    particle at position $\vb{Y}$, the modes of the field $\phi$ (red waves), and the particle--field interaction (blue box), which can store elastic energy $\Hint$. 
    Particle and field exchange heats $\dqy$ and $\dqp$, respectively, with a bath at temperature $T$.
    %
    An external agent exerts work $\dw$ on the particle. 
    Particle and field may exchange energies $\dwint_y$ and $\dwint_\p$, respectively, via their 
    interaction.
    For clarity we do not indicate work exchanges between field modes. 
    Heat fluxes are considered positive when they are supplied to the system.
    }
    \label{fig:setup}
\end{figure}

In this work we investigate the stochastic thermodynamics of a system consisting of a 
 mesoscopic, externally-driven probe
coupled to a
fluctuating 
medium. The latter is here represented by a
scalar field obeying non-conserved or conserved dynamics~\cite{halperin}, and
the whole system (probe$+$field) is immersed in a homogeneous 
heat bath at a fixed temperature ${{T}}$ that induces thermal fluctuations.
We provide suitable definitions for the heat, work and entropy 
exchanges between the probe, the field and the thermal bath, which are consistent with the first and second laws of stochastic thermodynamics.
This approach is simpler than addressing a fully microscopic model, while it goes 
beyond a description based on a generalized Langevin equation (GLE)~\cite{speck2007jarzynski,mai2007nonequilibrium,plati2022thermodynamic,di2020thermodynamic,puglisi2009irreversible,ohkuma2007fluctuation}, which may capture temporal, but not spatial correlations of the environment.
Notably, our framework is particularly powerful when dealing with a fluctuating medium that is close to a critical point, 
where one can 
replace a complex microscopic dynamics with
the simplest model 
belonging to
the same universality class~\cite{Stanley_1999,halperin}.
As a fruit of our approach, one can thereby investigate the interplay between the probe, the spatio-temporal correlated field, and the heat bath under the light of stochastic thermodynamics.
We illustrate the theory for the minimal yet insightful example of a
probe particle
dragged by a harmonic trap through a 
fluctuating Gaussian field.
This mimics the setup
commonly used to study friction and viscosity in (active)
microrheology experiments~\cite{ciliberto2017experiments,gomez2014probing,lintuvuori2010colloids},
and 
is prototypical in stochastic thermodynamics.

\smallskip
\textbf{The model. --- } 
We study a system formed by a mesoscopic probe at
position
$\vb{Y}(t)\in \mathbb{R}^d$ 
in $d$ spatial dimensions
and a scalar field, whose
value
at $\vb{x}\in \mathbb{R}^d$ and time $t$ is denoted by 
$\p(\vb{x},t)\in \mathbb{R}$. 
We assume $\phi$ to represent the relevant slow degrees of freedom 
of the medium, after 
integrating out all other possible faster modes.
For example, $\phi$ may represent the local magnetization in a spin system~\cite{halperin,demerypath}, the height 
of a fluctuating membrane over a reference plane~\cite{reister_lateral_2005,reister-gottfried_diffusing_2010,camley_contributions_2012,camley_fluctuating_2014}, the relative concentration of two chemical species in a near-critical binary liquid mixture~\cite{GambassiCCF}, etc., although in some of these cases the actual dynamics is more complex than the one considered here. 
%
The energy of the system is described by the 
Hamiltonian
\begin{equation}
    \mathcal{H}[\phi,\vb{Y},t] = \Hp[\phi]  + \Hint[\phi,\vb{Y}] +  \cor{U}(\vb{Y},t) ,
    \label{eq:fullH}
\end{equation}
where $\Hp$ denotes the 
energy of the field,
$\cor{U}$ is an external potential acting only on the probe, and
$\Hint$ encodes the interaction between the probe
and the field.
The probe and the scalar field are assumed to obey the following coupled overdamped Langevin equations:
\begin{align}
     \gamma_y \dot{\vb{Y}}  &= -  \grad_{\vb{Y}} \mathcal{H}
    +\vb{F}_\mathrm{ext} + \bm{\nu} , \label{eq:LEy} \\
    \gamma_\p \dot \phi &= -(-\nabla^2)^{\a}  \frac{\delta \mathcal H}{\delta \p} +\eta^{(\a)}, \label{eq:LEphi}
\end{align}
where $\vb{F}_\mathrm{ext}(\vb{Y},t)$ in \cref{eq:LEy} accounts for 
non-conservative 
external forces acting on the probe,
while in \cref{eq:LEphi} $\a=0$~or~$1$ for locally non-conserved or conserved field dynamics~\cite{halperin,Tauber}.
Moreover, \rev{$\gamma_{y,\p}$} are friction coefficients, while $\bm{\nu}$ and $\eta^{(\a)}$ are independent Gaussian white noises satisfying the
fluctuation-dissipation relations $\expval*{\nu_{i}(t) \nu_{j}(t') } = 2\gamma_y {{T}} \delta_{ij}\delta(t-t')$ and $\expval*{\eta^{(\a)}(\vb{x},t)\eta^{(\a)}(\vb{x}',t')}= 2\gamma_\phi  {{T}} (- \grad^2)^{\a}  \delta^d(\vb{x}-\vb{x}')\delta(t-t')$, where $\la\ldots \ra$ denotes the average over 
the
realizations of the noises 
(note that we set  Boltzmann's constant $k_B=1$ throughout the paper).
Accordingly, 
in the absence of external driving, i.e.,  $\vb{F}_\mathrm{ext}=0$ and time-independent $\cor{U}$, 
the system reaches a state of thermal equilibrium. 
Figure~\ref{fig:setup}(a) 
\rev{is} 
a sketch of the model 
\rev{with a harmonic external potential~$\mathcal U$.}

\smallskip
\textbf{First law of stochastic thermodynamics. --- }
\rev{To analyze the thermodynamic properties of this system, we generalize ideas from stochastic thermodynamics~\cite{sekimoto2010stochastic,seifert2012stochastic}. For the fluxes of energy and entropy between particle, field and thermal bath sketched in Fig.~\ref{fig:setup}(b), this yields the expressions below (derived in Ref.~\cite{SM}).}
First, 
the only systematic energy input into the 
system (probe+field) is the work exerted by an external agent on the probe. During a time interval $[t,t+\mathrm{d}t]$, this work is given by
\begin{equation}\label{def:w}
     \dw  = \frac{\partial \cor{U}}{\partial t} \dd t 
     + \vb{F}_\mathrm{ext}\circ \dd \vb{Y},
\end{equation}
where 
$\dbar$ denotes non-exact differentials, and $\circ$ indicates the usage of Stratonovich calculus throughout~\cite{seifert2012stochastic}. 
\revv{Requiring}
energy conservation for the probe 
leads
to the first law 
$\mathrm{d}\cor{U} = \dw + \dwint_y + \dqy$, according to which
any change $\mathrm{d}\cor{U}$ in
its potential energy
can either be due to $\dw$, or caused by the elastic coupling to the field, 
\begin{equation}\label{def:dwinty}
    \dwint_y :=-\grad_{\vb{Y}} \Hint \circ \mathrm{d}\vb{Y},
\end{equation}
\revv{or else}
dissipated into the bath in the form of heat $-\dqy$. From Eqs.~\eqref{eq:LEy}~to~\eqref{def:dwinty}
it directly follows that the latter is given by 
\begin{equation}
    \dqy = (-\gamma_y \dot{\vb{Y}}+\bm{\nu} )\circ \mathrm{d} \vb{Y} \label{def:qy},
\end{equation}
which is identical to 
the Sekimoto 
expression~\cite{sekimoto2010stochastic} for a 
single
particle in a heat bath (\ie, $\Hint\!=\!0$). 
Likewise, 
\revv{requiring}
local energy conservation for the field 
leads
to the first law for the field~\cite{SM} 
\begin{equation}
\label{eq:1stlawphi}
\mathrm{d} \Hp
=\int \dd[d]{\vb{x}}\left[  \dwint_\p (\vb{x})+\dqp (\vb{x}) \right].
\end{equation}
(An analogous, spatially-resolved first law is given in Ref.~\cite{SM}.)
Accordingly, all the work 
done on
$\p$ is either stored in the field configuration $\mathrm{d} \Hp$ (its ``bending''), or dissipated in the form of heat $\dbar Q_\p$.
Here, 
\begin{equation}
   \dwint_\p(\vb{x}) := -[{\delta \Hint}/{\delta \p (\vb{x})}] \circ \mathrm{d}\p(\vb{x}) 
\end{equation}
is the work locally \rev{exerted on} 
the field by the elastic coupling.
In general, $\dwint_y \neq -\int \dd[d]{\vb{x}}  \dwint_\p (\vb{x})$, because $\Hint$ itself can store 
energy.
From \cref{eq:1stlawphi,eq:LEphi} we 
find that the stochastic heat $\dqp$ takes the form
\begin{align}
    \dqp(\vb{x}) &= \left\lbrace (-\nabla^2)^{-\a} \left[\eta^{(\a)}(\vb{x})-\gamma_\p \dot \p(\vb{x})\right] \right\rbrace\circ \mathrm{d} \p(\vb{x}) \n\\
    &= \revv{\left[ \frac{\delta \Hp}{\delta \p(\vb{x}) } +\frac{\delta \Hint}{\delta \p(\vb{x}) } \right] \circ \mathrm{d} \p(\vb{x}),} \label{def:qphi}
\end{align}
generalizing Sekimoto's expression~\cite{seifert2012stochastic}.
Importantly, $\dqp$ is 
a density (or field) of local heat dissipation. For $\a=0$, \cref{def:qphi} formally resembles the heat exchange of a single particle.
The case $\a=1$ of conserved dynamics involves the inverse Laplace operator
$(\nabla^2)^{-1}$, 
which is nonlocal in space, and has thus to be interpreted in terms of its Green's function~\cite{bender}. In confined geometries, the latter depends on the boundary conditions~\cite{Gross_2021,Venturelli_2022_confined}, but here we will only consider the system in the bulk.


\smallskip
\textbf{Second law of stochastic thermodynamics. --- }
Next, we define the irreversibility~\cite{seifert2012stochastic,nardini2017entropy,Niggemann_2020} associated with an individual, joint (probe+field) trajectory $\{\vb{Y},\p \}_{\ti}^{\tf}$ as 
\begin{equation}\label{def:epr2}
     {S}_\mathrm{tot}
     =\ln \frac{\mathcal{P} [\{\vb{Y},\p \}_{\ti}^{\tf}]}{ \mathcal{ P}^\mathrm{R} [\{\vb{Y}^\mathrm{R},\p^\mathrm{R} \}_{\ti}^{\tf}]}.
\end{equation}
Here
$\mathcal{P}$ denotes the path probability of the trajectory, starting with the joint probability density 
$\rho_{\vb{Y},\p}[\vb{Y}({\ti}),\p(\vb{x},{\ti})]$, while $\mathcal{P}^\mathrm{R}$ denotes the path probability 
of the corresponding time-reversed process\footnote{We assume that $\p$ and $\vb Y$ are both even under time reversal. In the time-reversed process, the time-dependent external driving protocols $\cor{U}(\vb{x},t)+\vb{F}_\mathrm{ext}(t)$ are reversed in time. The backward trajectories are initialized with $\rho_{\vb{Y},\p}[\vb{Y}({\tf}),\p(\vb{x},{\tf})]$~\cite{SM}.
}.
By construction, the second law $\la  {S}_\mathrm{tot} \ra\geq 0$ holds~\cite{seifert2005entropy}.
A central 
result of this manuscript is that the irreversibility $ {S}_\mathrm{tot}$ defined in \cref{def:epr2} equals the stochastic (total) entropy production~\cite{SM}
\begin{equation}\label{def:epr}
     {S}_\mathrm{tot}= - \frac{Q_y}{T}-  \int \dd[d]{\vb{x}} \frac{Q_\p(\vb{x})}{T}  
    +\Delta S_{y,\p}^\mathrm{sh} ,
\end{equation} 
with the heat dissipation $Q_y$ and $Q_\p$ defined in \cref{def:qy,def:qphi}, respectively --- indicating the thermodynamic consistency of our framework\footnote{\revv{The definition of irreversibility given in \cref{def:epr2} generalizes those of, e.g., Ref.~\cite{nardini2017entropy} [see Eq.~(7) therein] and Ref.~\cite{Niggemann_2020} [see Eq.~(16) therein] in the presence of a probe. Similarly, \cref{def:epr} is compatible with the expression given in Eq.~(26) of Ref.~\cite{Seifert_2016} for 
a generic system strongly coupled to a heat bath, at the level of (total) spatially integrated quantities}.
}.
Here $\Delta S_{y,\p}^\mathrm{sh} = \ln \lbrace \rho_{\vb{Y},\p} [\vb{Y}(\ti),\p(\ti)]/ \rho_{\vb{Y},\p}[\vb{Y}(\tf),\p(\tf) ]\rbrace$
is the change in 
the fluctuating Shannon entropy.
The first and second terms 
on the~r.h.s.~of
Eq.~\eqref{def:epr} are the fluxes of entropy to the thermal bath from the probe and from the field, respectively.

If the system 
reaches a 
steady state, then
$\expval*{  \mathrm{d} \Hint }$, $\expval*{  \mathrm{d} \Hp }$, and $\expval*{ \mathrm{d} \cor{U}}$ vanish. 
This simplifies the first laws for probe and field, and implies $
\expval*{ \int\dd[d]{\vb{x}}\dwint_\p (\vb x)} = -\expval*{ \dwint_y }$. 
Together with
the second law and \cref{def:epr}, this implies 
\begin{equation}
    \label{eq:ssEPR}
    {\la \dot W \ra}= - \la \dot Q_y \ra -{ \int \dd[d]{\vb{x}}  \la \dot Q_\p(\vb{x})  \ra} 
    = T\la  \dot{S}_\mathrm{tot} \ra \geq 0,
\end{equation}
with $  \dot Q    :=  \dq /\mathrm{d}t $,
$  \dot W   :=  \dw /\mathrm{d}t $, and
$ \dot{S}_\mathrm{tot} \simeq  
 {S}_\mathrm{tot}/(\tf-\ti)
$. 
Thus, 
in \rev{such} steady states, $\expval*{ \dot{S}_\mathrm{tot}}$ is proportional to the average power $\langle \dot W\rangle$ injected into the system \revv{--- which is partly dissipated into heat directly by the probe, $\langle Q_y\rangle $, and partly through the field, $\langle Q_\phi \rangle$.}

\begin{figure}
    \centering
    \includegraphics[width=\columnwidth]{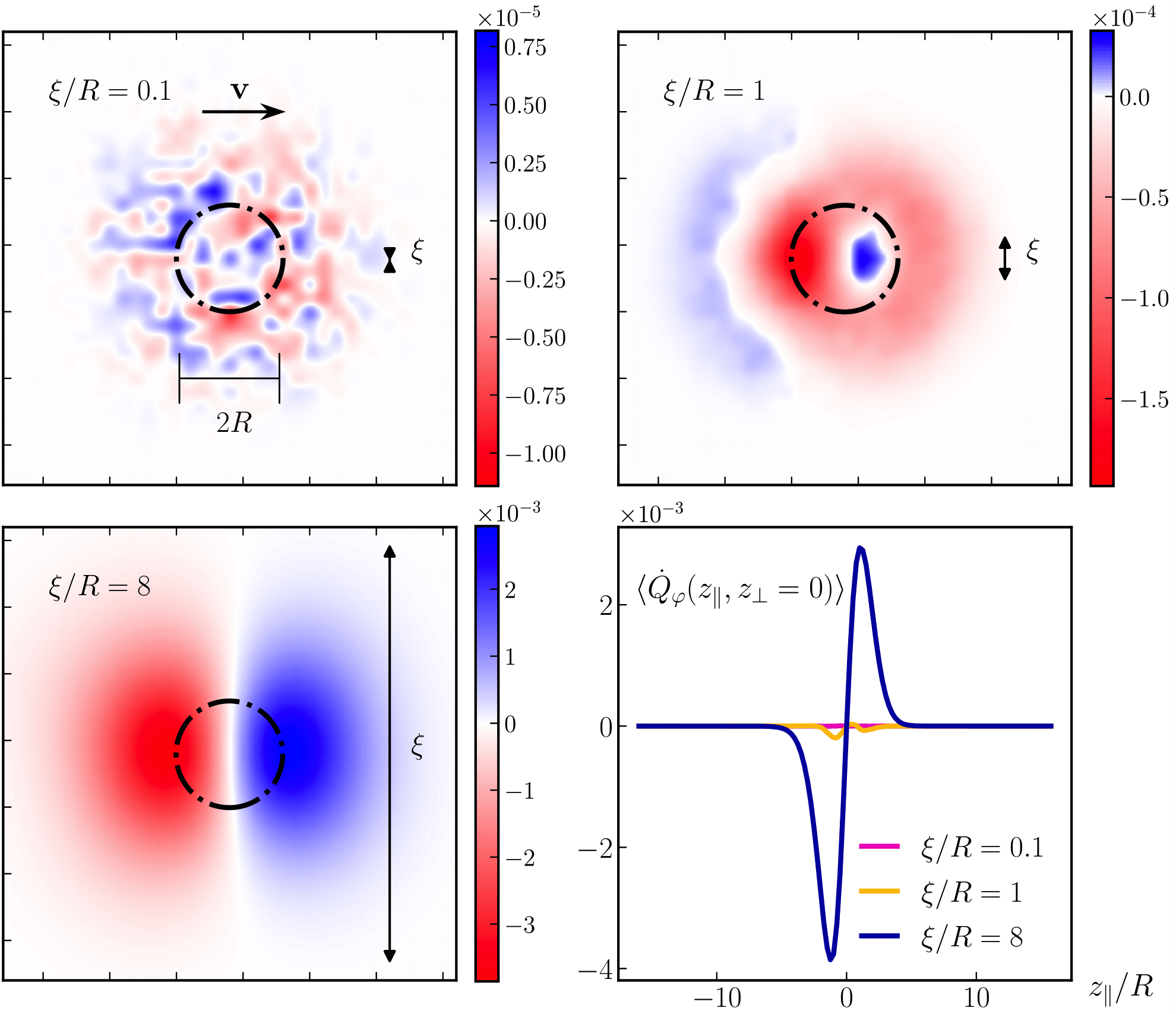}
    \put(-247,118){\footnotesize (a)}
    \put(-115,118){\footnotesize (b)}
    \put(-247,12){\footnotesize (c)}
    \put(-115,12){\footnotesize (d)}
    \put(-246,30){\scriptsize \textcolor{red}{Heat}}
    \put(-246,22){\scriptsize \textcolor{red}{dissipation}}
    \put(-190,22){\scriptsize \textcolor{blue}{Heat}}
    \put(-210,14){\scriptsize \textcolor{blue}{absorption}}
    \put(-247,160){\includegraphics[scale=0.15]{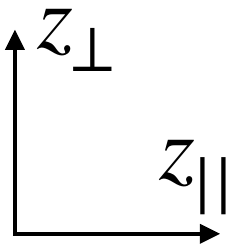}}
    \caption{ 
    Average local heat dissipation rate $\expval*{\dot Q_\varphi(\vb z)}$ by the field into the bath, 
    in a frame comoving at velocity $\vb v$ with the harmonic trap (see \cref{fig:setup}).
    \rev{These results of numerical simulations~\cite{SM} refer to}
    a Gaussian field in $d=2$ with 
    non-conserved dynamics
    \rev{and, from (a) to (c),} increasing values of $\xi/R$, where $\xi$ is the correlation length and $R$ the probe size.
    \rev{The dash-dotted circles indicate the size of the probe.}
    \rev{Panel (d) shows}
    $\expval*{\dot Q_\varphi(\vb z)}$ along the drag direction $z_\parallel$, for $z_\perp=0$ and the values of $\xi$ considered in (a--c).  
    We used $\lambda=5$, $v=5$, \revv{lattice size} $L=128$, \revv{time step} $\Delta t = 10^{-2}$, a Gaussian potential $V$ with variance $R=4$,  
    and we set all other parameters to unity.
    }
    \label{fig:heat_field}
\end{figure}
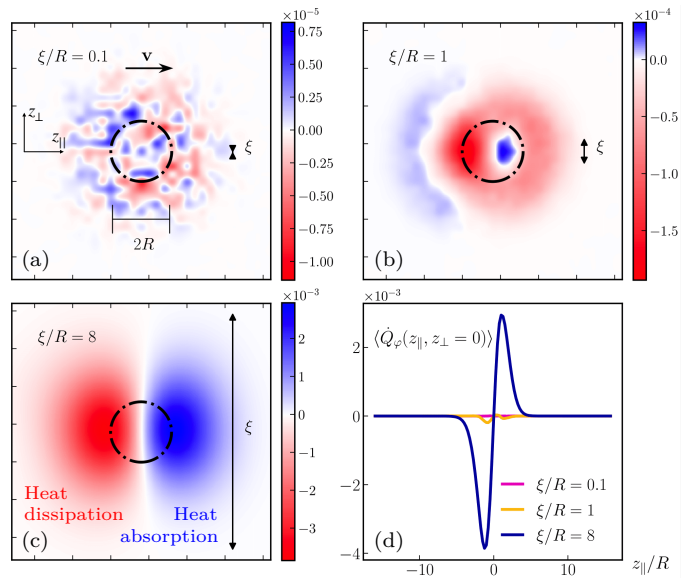

\smallskip
\textbf{Example: Particle dragged through a Gaussian field. --- }
Within
this framework, we 
consider
the typical
setup~\cite{ciliberto2017experiments} in which a probe particle is 
confined
in a harmonic trap of stiffness $\kappa$ moving at constant velocity $\mathbf{v}$.
The corresponding potential is
$ \cor{U}(\vb{Y},t)=\kappa\left(\vb{Y} - \mathbf{v} t\right)^2/2 $,
and the dissipation rate
follows from \cref{def:w} as
\begin{equation}
\dot W  =
    -\kappa \vb{v}\cdot \left( \vb{Y}-\vb{v} t \right) .
    \label{eq:def_power}
\end{equation}
In the long-time limit, the system reaches a steady state in the comoving reference frame with velocity $\vb{v}$. 
As a 
minimal model for a near-critical medium --- and as the simplest approximation of various complex systems \revv{displaying long-range spatial correlations} --- we consider a Gaussian field with Hamiltonian~\cite{halperin}
\begin{equation}
        \cor{H}_\phi= \frac{1}{2}\int \dd[d]{\vb{x}}
        \left[ (\nabla\phi)^2+r\phi^2\right].
        \label{eq:gaussian_hamiltonian}
\end{equation}
The correlation length $\xi=r^{-1/2}\geq 0$ controls the spatial range of the field correlations at equilibrium, and diverges upon approaching the critical point $r = 0$.
We model the particle--field 
interaction, 
as in 
\ccite{demery2010, demery2010-2, demery2011, demerypath, Dean_2011, demery2013}, 
by
\begin{equation}
    \cor{H}_\T{int} = -\lambda \int \dd[d]{\vb{x}} \phi(\vb{x})V(\vb{x}-\vb{Y}),
    \label{eq:Hint}
\end{equation}
where $V(\vb{x})\geq 0$ is 
a rotationally invariant function, \rev{with unit spatial integral and
rapidly vanishing as $|\vb x|$ increases beyond the ``radius'' $R$ of the particle,} 
while $\lambda$ sets the coupling strength. 
For $\lambda >0$ and at equilibrium, configurations in which the field is enhanced
around
the particle are energetically favored.
We express the particle position in the 
comoving frame as $\vb{Z}\equiv \vb{Y}-\vb{v}t +\vb{v}\gamma_y/\kappa$,
so
that $\vb{Z}=0$ is the 
mean steady-state position for $\lambda=0$. Similarly, we introduce the 
field $\varphi(\vb{z},t)\equiv \phi(\vb{z}+\vb{v}t -\vb{v}\gamma_y/\kappa,t)$ in the comoving frame.

\begin{figure*}
\centering
\includegraphics[width=\columnwidth]{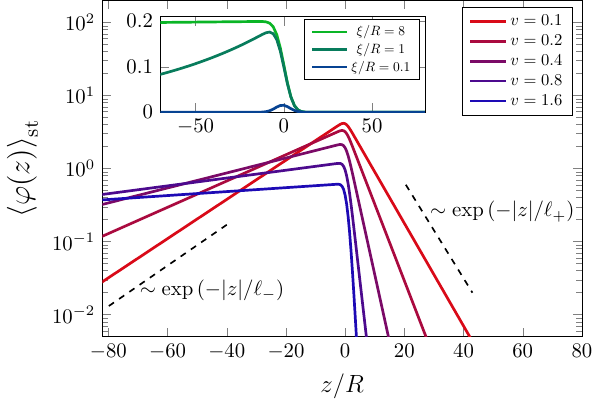}
\includegraphics[width=\columnwidth]{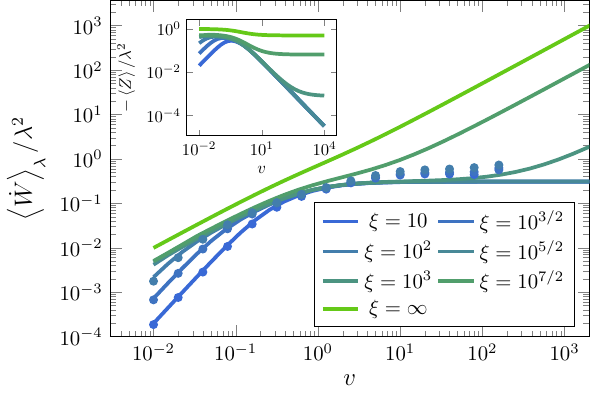}
  \put(-501,160){(a)}
  \put(-246,160){(b)}
\caption{\textbf{(a)} 
    Steady-state expectation value $\langle \varphi(z) \rangle_{\rm st}$ of the field in the comoving frame, 
    for various drag velocities $v>0$, in $d=1$ and with non-conserved dynamics.
    The \textit{shadow} $\langle \varphi(z) \rangle_{\rm st}$ flattens upon increasing $v$ (at fixed correlation length $\xi=10$, main plot) or upon decreasing $\xi$ (at fixed $v=5$, inset). 
    $\langle \varphi(z) \rangle_{\rm st}$ decays exponentially upon increasing $|z|$ with different decay lengths 
    $\ell_\pm$
    in front of or behind the particle --- see the main text and Ref.~\cite{SM}.
    We used
    $V_q=\exp(- q^2 R^2/2)$ and $R^2 + {{T}}/\kappa = 1$, 
    while the other parameters were set to unity.
    \textbf{(b)}
    \rev{Additional} dissipation rate [see \cref{eq:correction_power}]
    as a function of $v$,
    for various values of $\xi$ 
    \rev{(with $T=0.1$, while the other parameters were set to one).}
    Symbols correspond to simulations. 
    Inset: scaled correction $-\avg{Z}/\lambda^2$ to the average particle position 
    [see Eq.~\eqref{eq:avg_position}]. 
    } 
\label{fig:regimes}
\end{figure*}

\smallskip
\textbf{Heat dissipation field. --- } 
The first thermodynamic quantity we analyze is the 
spatially-resolved heat dissipation $Q_\varphi$ in the comoving reference frame~\cite{SM}. In Fig.~\ref{fig:heat_field} we show numerical results for $\la \dot Q_\varphi \ra$ of a field in $d=2$ with $\a=0$. 
For 
small values of $\xi$ [see panel (a)], 
we observe that 
$\la \dot Q_\varphi \ra$
is essentially negligible (within numerical uncertainties) and displays no discernible spatial structure.
In contrast, if $\xi$ reaches the order of the particle size $R$ [panel (b)], 
regions of average heat dissipation ($\la{\dot Q_\varphi}\ra <0$) or absorption  ($\la{\dot Q_\varphi}\ra >0$) start developing. 
Close to criticality, with $\xi \gg R$ [panel (c)], a 
dissipation dipole 
forms, with a region of heat absorption in front of the particle, whose spatial extent is approximately given by $\xi$. Hence, surprisingly,
in front of the particle the heat bath supplies net energy as if it was coupled to a cooler object. Note that the second law implies $\la{\dot Q_y}\ra+\int \dd^d{\vb x} \la{\dot Q_\varphi(\vb x) }\ra < 0  $, but it does not preclude local heat absorption, i.e., $\la{\dot Q_\varphi(\vb x) }\ra > 0$ for some $\vb x$. 
To further elucidate the origin of this effect, below we analytically 
investigate 
the statistics of particle and field for various values of $v=|\vb v|$ and $\xi$, and 
the dissipated power.

\smallskip
\textbf{Particle statistics and bending of the field. --- } 
We now assume $\lambda$ to be small, and use it as a perturbative parameter~\cite{Venturelli_2022, wellGauss,Venturelli_2022_2parts,Gross_2021,Venturelli_2023}.
Expressing
\cref{eq:LEy,eq:LEphi} in terms of $\vb{Z}$ and $\varphi$, we first obtain
\begin{align}
    &\gamma_y \dot{\vb{Z}} = -\kappa \vb{Z} +\lambda \int \dslash{p} i \vb{p}\, V_{-p}\, \varphi_{\vb p}\,  e^{i \vb{p}\cdot \vb{Z}} + \bm{\nu} , \label{eq:frame_z_main} \\
    &\gamma_\phi (\partial_t - i\vb{p}\cdot \vb{v} +\alpha_p) \varphi_{\vb p} 
    = \lambda p^{2\a} V_p e^{-i \vb{p}\cdot \vb{Z}} + \eta_{\vb p}^{(\a)} , 
    \label{eq:frame_varphi_main} 
\end{align}
where 
$\alpha_p \equiv p^{2\a}(p^2+r)/\gamma_\phi$, and $\varphi_{\vb p}$ is the Fourier transform\footnote{We adopt here and in the following the Fourier convention $f(\vb{x}) = \int [\dd[d]{p}/(2\pi)^d] \exp(i\vb{p}\cdot \vb{x}) f_{\vb p}$, and we normalize the delta distribution in Fourier space as $\int [\dd[d]{p}/(2\pi)^d] \delta^d(\vb p)=1$.} of $\varphi(\vb x,t)$.
%
From \cref{eq:frame_varphi_main} 
we can 
\rev{calculate the} mean stationary profile $\avg{\varphi_{\vb p}}_\T{st}$,
which is comoving with the particle, and which we call ``shadow''~\cite{SM}.
The latter 
is shown in Fig.~\ref{fig:regimes}(a) for a field 
\rev{in $d=1$ with $\a=0$:}
the field is strongly bent around the particle, while $\expval*{ \varphi(z)}_\T{st} \propto\exp(-|z|/\ell_\pm)$ for 
\rev{$z \to \pm \infty$, with $\ell_\pm = \xi [ \sqrt{1+(\xi v\gamma_\phi/2)^2}  \mp (\xi v\gamma_\phi/2) ]$~\cite{SM}.}
Far from criticality ($\xi\to 0$, see inset), the shadow 
\rev{vanishes,}
rationalizing the \rev{corresponding vanishing of}
$\expval*{\dot Q_\varphi(\vb z)}$ in \cref{fig:heat_field}(a).

Via a perturbative approach, we can \rev{investigate} 
the particle fluctuations analytically. The moment generating function $g(\vb{q})\equiv \expval*{\exp(-i\vb{q}\cdot \vb{Z})}$ of the particle position at the lowest nontrivial order in 
$\lambda$
reads~\cite{SM}
\begin{align}
    &\ln g(\vb{q}) =  -\frac{q^2 {{T}}}{2\kappa}+\frac{i \lambda^2}{\kappa} \int_0^\infty \!\frac{\dd{u}}{\sigma^2(u)} \int \!\dslash{p} \frac{\left( \vb{p}\cdot \vb{v} \right)}{p^2+r} |V_p|^2 \n\\
    &\quad\times  G_{\vb p}(u)e^{-p^2\sigma^2(u)}\left[1-  e^{-\left( \vb{p}\cdot \vb{q} \right) \sigma^2(u) } \right] +\order{\lambda^4}, \label{eq:position_cgf}
\end{align}
where $
    \sigma^2(u) \equiv {{T}}\left( 1-e^{-\kappa u/\gamma_y} \right)/\kappa$,
and $
    G_{\vb p}(u) \equiv \exp[-\alpha_p u +i \left( \vb{p}\cdot \vb{v} \right) u ] $.
\rev{Interestingly, \cref{eq:position_cgf} predicts a non-Gaussian statistics
of the particle position. In addition,}
the variance $\expval*{Z_l Z_m}_c\propto \delta_{lm}$ 
\rev{changes}
anisotropically compared 
to the 
case $\vb v= \vb 0$, 
so that the position 
distribution is
elongated 
in the direction parallel 
to $\vb{v}$~\cite{Demery_2019,SM}.
Furthermore, $i\nabla_{\vb{q}}g|_{\vb{q}=\vb{0}}$ 
gives the average displacement
\begin{equation}
    \expval*{\vb{Z}} = - \frac{\lambda^2}{\kappa} \int\! \dslash{p} \frac{\vb{p} \left( \vb{p}\cdot \vb{v} \right)}{p^2+r}  |V_p|^2 \int_0^\infty\! \dd{u} G_{\vb p}(u) e^{-p^2\sigma^2(u)}, \label{eq:avg_position}
\end{equation}
which turns out to be directed along $-\vb{v}$; hence, the field induces a shift of the probability density of the particle position, which 
lags behind the average stationary value in the absence of the field [see the inset of Fig.~\ref{fig:regimes}(b)]. Such a lag is the footprint of an underlying \rev{additional} 
source of dissipation, which we analyze next. 

\smallskip
\textbf{Power fluctuations. --- } 
From the moment generating function \rev{in \cref{eq:position_cgf},} 
we can access the distribution of the dissipated power.
Indeed, rewriting \cref{eq:def_power} in terms of $\vb{Z}$
gives 
$\dot{W} = \gamma_y v^2 - \kappa \vb{v} \cdot {\vb{Z}}$,
and thus
\begin{equation}
\ln \expval{\exp(- i\mu \dot{W})} = -i\mu \gamma_y v^2 +\ln g(-\mu \kappa \vb{v}),
\end{equation}
encoding all moments of the dissipated power.
To study the impact of the field, \rev{we focus on} 
the average power 
\begin{equation}
    \expval*{\dot{W}} = \gamma_y v^2 - \kappa \vb{v} \cdot \expval*{\vb{Z}} \equiv \expval*{\dot W}_{0} + \expval*{\dot W}_{\lambda} ,
    \label{eq:correction_power}
\end{equation}
where we identify the dissipation rate in the absence of the field $\expval*{\dot W}_{0}= \gamma_y v^2\geq 0 $, 
while $
\expval*{\dot W}_{\lambda} = - \kappa \vb{v} \cdot \avg{\vb{Z}} \geq 0$ 
encodes the \rev{additional} 
dissipation due to the field.
According to Eq.~\eqref{eq:ssEPR}, these are further equal to the entropy production at $\lambda=0$ and $\lambda>0$, i.e., $\expval*{\dot W}_{0}=T\la  \dot S_\mathrm{tot} \ra_0$ and $\expval*{\dot W}_{\lambda}= T\la  \dot S_\mathrm{tot} \ra_\lambda$, \rev{respectively.} 
In Fig.~\ref{fig:regimes}(b) 
we show 
the perturbative prediction for
$\expval*{\dot W}_{\lambda}$
as a function of 
$v$, 
and the corresponding numerical data from simulations~\cite{SM}, which are in good agreement.
We find that $\expval*{\dot W}_{\lambda}$ generally displays 
three distinct regimes \rev{upon increasing $v$}: first, $\abs{\la Z \ra}$ grows linearly 
[see Fig.~\ref{fig:regimes}(b), inset], so that 
$\expval*{\dot{W}}_\lambda \propto v^2$, as it would be the case 
for usual Stokes friction.
\rev{After a crossover,} in the second regime $\abs{\la Z \ra} \propto 1/v$,
\rev{and therefore}
$\expval*{\dot{W}}_\lambda$
plateaus at intermediate $v$,
which indicates a constant energetic cost associated with the particle--field interaction.
Finally, \rev{in the third regime}, $\abs{\la Z \ra}$ saturates and thus $\expval*{\dot{W}}_\lambda\propto v$.
We remark that \rev{the second and third} 
\textit{non-Stokesian} regimes\footnote{The 
first and the third among these regimes
are consistent with the scaling of drag forces reported in Refs.~\cite{demery2010,demery2010-2,demery2011}, where particles moving with constant velocity (i.e., without positional fluctuations) were studied.} cannot be captured by a linear GLE~\cite{SM}.

We present a thorough analysis of these regimes in \ccite{SM}, \rev{for the case $\a=0$,}
and summarize it here.
First, 
by inserting into \cref{eq:frame_z_main} the formal solution of \cref{eq:frame_varphi_main} for $\varphi_{\vb p}(t)$,
an effective equation 
for
$\vb{Z}(t)$ can be obtained, which is non-Markovian and nonlinear.
However, in the limit of 
small $v$, the field is sufficiently fast to equilibrate around the 
particle at any instant in time~\cite{Venturelli_2022_2parts,Gross_2021}. Conversely, for 
large $v$ the evolution of the field is so slow that the particle 
encounters 
an effectively static field configuration.
Accordingly, the first two regimes can be
quantitatively captured 
by adiabatically
replacing the field $\varphi_{\vb p}(t)$ in \cref{eq:frame_z_main} with its mean (comoving) profile $\avg{\varphi_{\vb p}}_\T{st}$, i.e., with the {shadow} shown in
Fig.~\ref{fig:regimes}(a), resulting in an approximately Markovian evolution of $\vb{Z}(t)$.
\rev{In contrast, at intermediate values of $v$, 
the timescales  of the particle dynamics 
are comparable with the relaxation time 
$\tau_\xi \simeq \gamma_\phi \xi^2$ 
of the field, and the adiabatic approximation is no longer accurate:} 
the particle 
dynamics
\rev{within the crossover between the first two regimes}
is dominated by the memory effects caused by the mutual influence of the particle and the field~\cite{SM}.
\rev{Finally},
in the third regime, 
the shadow 
becomes negligible compared to the (critical) fluctuations of the field, and the particle effectively encounters a rough landscape resulting from them. 
Notably, as the field approaches criticality ($\xi\to \infty$), the amplitude of its fluctuations diverges~\cite{Tauber}, and thus the last (non-Stokesian) regime extends to low $v$.

\smallskip
\textbf{Conclusions. --- }
We 
developed
a thermodynamically consistent framework to study the 
energetic and entropic flows 
for a probe in a fluctuating 
medium with spatio-temporal correlations, 
modeled here by a scalar field
immersed in a heat bath.
We showed that the mutual influence 
\rev{of the probe and the correlated environment}
leads to unusual
thermodynamic properties, even for the simple 
example of a particle dragged by a harmonic trap through a Gaussian field. 
We 
\rev{showed}
that, close to criticality or, more generally, when the correlation length $\xi$ of the field exceeds the particle size, a dipolar structure develops in the local heat dissipation field --- with systematic heat absorption in front of the particle,
and whose extent
is \rev{determined} by~$\xi$. We thus expect such {\em heat dipoles} to be possibly revealed 
also in
long-range interacting systems of diverse nature~\cite{gupta2017world}.
Furthermore, the 
\rev{additional} power required to drag the particle 
in the presence of
the field features three regimes 
with distinct scaling in the drag speed $v$, 
among which are
two 
non-Stokesian regimes --- a feature that cannot be captured by a linear 
GLE.
Far from criticality ($\xi\to 0$), 
the medium is only weakly correlated, 
and indeed both the heat dipole and the 
\rev{additional} dissipation vanish. 
The framework 
developed
here can be readily applied to study more complex scenarios, e.g., non-quadratic Hamiltonians, or extended to systems of multiple particles in a common correlated (active) environment. 
Moreover, 
the recent and intense experimental investigations of colloidal particles in correlated fluids --- such as viscoelastic media~\cite{gomez2014probing,ginot2023average,loos2023universal}, binary liquid mixtures~\cite{Hertlein_2008,Gambassi_2009,volpe2011microswimmers,Paladugu_2016,Ciliberto_2017,Magazzu_2019,martinez2023laserinduced}, and living cells~\cite{fodor2016nonequilibrium,turlier2016equilibrium} --- could 
provide access to the phenomena described here.

\acknowledgments
SL acknowledges funding through the Walter Benjamin program by the Deutsche Forschungsgemeinschaft (German Research Foundation, project number~498288081), and through the postdoctoral fellowship by the Marie Skłodowska-Curie Actions (MSCA), undertaken by the UKRI (grant reference~EP/X031926/1).
AG, DV, and BW acknowledge support from MIUR~PRIN project “Coarse-grained description for non-equilibrium systems and transport phenomena (CO-NEST)” n.~201798CZL. BW further acknowledges funding from the Imperial College Borland Research Fellowship. ER and AG acknowledge financial support from
PNRR~MUR project PE0000023-NQSTI.

\bibliographystyle{eplbib}
\bibliography{bibliography}


\clearpage


\section*{Supplementary material}

# Supplemental Material
# for "Stochastic thermodynamics of a probe in a fluctuating correlated field"

Davide Venturelli,[1,2,*] Sarah A. M. Loos,[3,4,*] Benjamin Walter,[5,1,*] Édgar Roldán,[4] and Andrea Gambassi[1]

[1]*SISSA — International School for Advanced Studies and INFN, via Bonomea 265, 34136 Trieste, Italy*
[2]*Laboratoire de Physique Théorique de la Matière Condensée,
CNRS/Sorbonne Université, 75005 Paris, France*
[3]*DAMTP, Centre for Mathematical Sciences, University of Cambridge,
Wilberforce Rd, CB3 0WA Cambridge, United Kingdom*
[4]*ICTP — International Centre for Theoretical Physics, Strada Costiera 11, 34151 Trieste, Italy*
[5]*Department of Mathematics, Imperial College London,
180 Queen's Gate, SW7 2AZ London, United Kingdom*
(Dated: April 30, 2024)

In this Supplemental Material we provide additional analytical derivations and numerical results which support the discussion in the main text, and justify some of the results stated therein.

## CONTENTS

## I. STOCHASTIC ENERGETICS FOR THE PARTICLE AND FIELD

In order to address the thermodynamic properties of the system described in our manuscript — composed of a probe and a field in interaction — we need to introduce suitable definitions of the energy flows and entropy changes associated with their stochastic dynamics. To this end, we utilize ideas from the framework of stochastic thermodynamics [1], and generalize them to the present case. In addition to the equations of motion, given in Eqs. (2) and (3) in the main text, the only assumption we make is *energy conservation* for both the probe and field dynamics. We use Stratonovich calculus throughout.

First, we assume that the only systematic energy input into the system is due to the work done on the probe by the action of an external agent. This can either result from changing in time the potential $\mathcal{U}(\mathbf{Y},t)$ — thereby increasing

the potential energy of the probe — or from the application of an external force $\mathbf{F}_{\text{ext}}$. In agreement with classical mechanics, the work $đW$ within the time frame $[t, t + \mathrm{d}t]$ is simply given by

$$đW = \frac{\partial \mathcal{U}(\mathbf{Y}, t)}{\partial t} \mathrm{d}t + \mathbf{F}_{\text{ext}} \circ \mathrm{d}\mathbf{Y}. \tag{1}$$

The total work $W$ along a certain stochastic trajectory $\{\mathbf{Y}(t), \phi(\mathbf{x}, t)\}_{t_\text{i}}^{t_\text{f}}$ in the configuration space of the field and probe, from time $t_\text{i}$ to $t_\text{f}$, is obtained by integrating over the infinitesimal increments, i.e., $W[\{\mathbf{Y}, \phi\}_{t_\text{i}}^{t_\text{f}}] = \int_{\{\mathbf{Y}, \phi\}_{t_\text{i}}^{t_\text{f}}} đW$.

Next, we consider the total energy $\mathcal{U} + \mathcal{H}_\phi + \mathcal{H}^{\text{int}}$ of the system and how it may change, in order to identify all possible types of energy (ex-)changes within the system. First, the probe can store potential energy $\mathcal{U}(\mathbf{Y}, t)$. The latter can change due to a probe displacement, i.e., a change of $\mathbf{Y}$, or due to the motion of the trap itself, giving rise to the differential

$$\mathrm{d}\mathcal{U}(\mathbf{Y}, t) = \boldsymbol{\nabla}_\mathbf{Y} \mathcal{U}(\mathbf{Y}, t) \circ \mathrm{d}\mathbf{Y} + \frac{\partial \mathcal{U}(\mathbf{Y}, t)}{\partial t} \mathrm{d}t. \tag{2}$$

Note that the last term on the r.h.s. of this equation can be expressed, via Eq. (1), in terms of the work $đW$ as $đW - \mathbf{F}_{\text{ext}} \circ \mathrm{d}\mathbf{Y}$. This fact will be used further below in Sec. IA. Second, the field can store energy $\mathcal{H}_\phi$ in its configuration, characterized by its local energy density $h_\phi(\mathbf{x})$, such that

$$\mathcal{H}_\phi = \int \mathrm{d}^d\mathbf{x}\, h_\phi(\mathbf{x}). \tag{3}$$

Since we assume that $\mathcal{H}_\phi$ is not explicitly time-dependent, the total change $\mathrm{d}\mathcal{H}_\phi$ of the internal energy is accordingly given by

$$\mathrm{d}\mathcal{H}_\phi = \int \mathrm{d}^d\mathbf{x}\, \mathrm{d}h_\phi(\mathbf{x}), \tag{4}$$

with

$$\mathrm{d}h_\phi(\mathbf{x}) = \frac{\delta \mathcal{H}_\phi}{\delta \phi(\mathbf{x})} \circ \mathrm{d}\phi(\mathbf{x}). \tag{5}$$

Here $\mathrm{d}h_\phi(\mathbf{x})$ denotes the local change of internal energy of the field. Notably, depending on the choice of $\mathcal{H}_\phi$, $\mathrm{d}h_\phi(\mathbf{x})$ may contain contributions from the neighboring points $\phi(\mathbf{x}' \neq \mathbf{x})$. For example, for a Gaussian Hamiltonian $\mathcal{H}_\phi = \int \mathrm{d}^d\mathbf{x} \left[\frac{1}{2}(\nabla\phi)^2 + \frac{1}{2}r\phi^2\right]$, as given in Eq. (14) in the main text, the Laplacian term induces energy exchanges between neighboring locations, as one realizes by discretizing space on a lattice.

Third, the interaction $\mathcal{H}^{\text{int}}$ between field and probe can store elastic energy. Again, defining the density of interaction energy $h^{\text{int}}$ such that

$$\mathcal{H}^{\text{int}} = \int \mathrm{d}^d\mathbf{x}\, h^{\text{int}}(\mathbf{x}), \tag{6}$$

the differential of the latter reads

$$\mathrm{d}\mathcal{H}^{\text{int}}(\mathbf{x}, t) = -đW_y^{\text{int}} - \int \mathrm{d}^d\mathbf{x}\, đW_\phi^{\text{int}}(\mathbf{x}), \tag{7}$$

hence

$$đW_y^{\text{int}} := -\boldsymbol{\nabla}_\mathbf{Y} \mathcal{H}^{\text{int}} \circ \mathrm{d}\mathbf{Y}, \tag{8}$$

$$đW_\phi^{\text{int}}(\mathbf{x}) := -\frac{\delta \mathcal{H}^{\text{int}}}{\delta \phi} \circ \mathrm{d}\phi. \tag{9}$$

Here, we have defined the scalar $đW_y^{\text{int}}$ and $đW_\phi^{\text{int}}(\mathbf{x})$ (the latter being a field) as the infinitesimal changes of the interaction energy due to the fluctuations of $\mathbf{Y}$ or $\phi(\mathbf{x})$, respectively. We interpret these energy flows as the work done by $\mathbf{Y}$ and $\phi$, respectively.

### A. First law of thermodynamics

As a final step, we connect the various energy flows, thereby obtaining the balance equations and the appropriate expressions for the heat. We start with $d\mathcal{U}(\mathbf{Y},t)$ given in Eq. (2): using Eq. (1), one finds

$$d\mathcal{U}(\mathbf{Y},t) = [\boldsymbol{\nabla}_\mathbf{Y}\mathcal{U}(\mathbf{Y},t) - \mathbf{F}_{\text{ext}}] \circ d\mathbf{Y} + \text{d}W. \tag{10}$$

Substituting the equation of motion of the probe dynamics [given in Eq. (2) of the main text] into the last equation, we find

$$\begin{aligned} d\mathcal{U}(\mathbf{Y},t) &= \left(\boldsymbol{\nu} - \gamma_y \dot{\mathbf{Y}} + \boldsymbol{\nabla}_\mathbf{Y}\mathcal{H}^{\text{int}}\right) \circ d\mathbf{Y} + \text{d}W \\ &= \left(\boldsymbol{\nu} - \gamma_y \dot{\mathbf{Y}}\right) \circ d\mathbf{Y} + \text{d}W_y^{\text{int}} + \text{d}W. \end{aligned} \tag{11}$$

In the last step we have identified $\text{d}W_y^{\text{int}}$, given by Eq. (8). Since we assume energy conservation for the probe dynamics, all the work $\text{d}W_y^{\text{int}} + \text{d}W$ done on the system that is not used to increase the potential energy of the probe (by an amount $d\mathcal{U}$) must be dissipated into the heat bath in the form of the heat $-\text{d}Q_y$. Thus, it follows from Eq. (11) that the first law for the probe dynamics reads

$$d\mathcal{U} = \text{d}Q_y + \text{d}W_y^{\text{int}} + \text{d}W, \tag{12}$$

with the heat flow $\text{d}Q_y$ defined by

$$\text{d}Q_y = (\boldsymbol{\nu} - \gamma_y \dot{\mathbf{Y}}) \circ d\mathbf{Y}, \tag{13}$$

as given in Eq. (6) of the main text.

We now turn to the energetics of the field. We start with $dh_\phi(\mathbf{x})$ given in Eq. (5) and we insert the equation of motion of the field [given in Eq. (3) of the main text], finding

$$dh_\phi(\mathbf{x}) = \frac{\delta \mathcal{H}_\phi}{\delta \phi(\mathbf{x})} \circ d\phi(\mathbf{x}) = \left\{ (-\nabla^{-2})^{\mathfrak{a}} \left[ \eta^{(\mathfrak{a})}(\mathbf{x}) - \gamma_\phi \dot{\phi}(\mathbf{x}) \right] + \frac{\delta \mathcal{H}^{\text{int}}}{\delta \phi} \right\} \circ d\phi(\mathbf{x}). \tag{14}$$

Identifying the last contribution on the r.h.s. of this equation with $\text{d}W_\phi^{\text{int}}$ [see Eq. (9)], we find

$$dh_\phi(\mathbf{x}) = \left\{ (-\nabla^{-2})^{\mathfrak{a}} \left[ \eta^{(\mathfrak{a})}(\mathbf{x}) - \gamma_\phi \dot{\phi}(\mathbf{x}) \right] \right\} \circ d\phi(\mathbf{x}) + \text{d}W_\phi^{\text{int}}(\mathbf{x}). \tag{15}$$

Again, energy conservation requires that all the work $\text{d}W_\phi^{\text{int}}$ that is locally done on the field is either stored in the field configuration in the form of a $dh_\phi(\mathbf{x})$, or dissipated by the field in the form of heat $-\text{d}Q_\phi(\mathbf{x},t)$. Accordingly, the (local) first law for the field, valid at any point in space, reads

$$dh_\phi(\mathbf{x}) = \text{d}Q_\phi(\mathbf{x}) + \text{d}W_\phi^{\text{int}}(\mathbf{x}). \tag{16}$$

We note that the l.h.s. describes the total change of internal energy of the field at a given point $\mathbf{x}$, which generally also contains nonlocal contributions, as mentioned after Eq. (5). These terms can be interpreted as the energy exchanges within the field, i.e., between $\phi$ at different points in space. The integral form of Eq. (16) is given in Eq. (7) in the main text. This first law allows us to identify from Eq. (15) the heat dissipation field of $\phi$ as

$$\text{d}Q_\phi(\mathbf{x}) = \left\{ (-\nabla^{-2})^{\mathfrak{a}} \left[ \eta^{(\mathfrak{a})}(\mathbf{x}) - \gamma_\phi \dot{\phi}(\mathbf{x}) \right] \right\} \circ d\phi(\mathbf{x}), \tag{17}$$

which is reported in Eq. (9) in the main text.

The accumulated heat and work along a joint trajectory $\{\mathbf{Y}, \phi\}_{t_i}^{t_f}$ from $t_i$ to $t_f$ are obtained by integrating over the infinitesimal increments defined above, i.e.,

$$e[\{\mathbf{Y}, \phi\}_{t_i}^{t_f}] = \int_{\{\mathbf{Y}, \phi\}_{t_i}^{t_f}} \text{d}e, \tag{18}$$

with $e \in \{Q_y, W, \text{d}Q_\phi(\mathbf{x})\}$.

Explicit expressions of $\langle \dot{Q}_\phi(\mathbf{x}) \rangle$ for a scalar Gaussian field coupled to a dragged particle [see Eqs. (14) and (15) in the main text] will be reported in a future work.

### B. Long-time limit

For $t \to \infty$, the system approaches a state in which the probability densities of the probe's degree of freedom $\mathbf{Y}$ and the field value $\phi$ in the comoving frame of reference [e.g., $\mathbf{x} \to \mathbf{x} - \mathbf{v}t$ w.r.t. the lab frame for a uniformly dragged particle] are time-independent. In this nonequilibrium steady state, the thermodynamic laws simplify due to the presence of additional constraints. In particular, the potential energy of the probe and of the field, as well as the interaction Hamiltonian, are on average conserved, such that $\langle \mathrm{d}\mathcal{U} \rangle = 0$, $\langle \mathrm{d}h_\phi \rangle = 0$, and $\langle \mathrm{d}h^{\mathrm{int}} \rangle = 0$. Thus, the first laws for the probe and field given in the main text and reported here in Eqs. (12) and (16), respectively, simplify to

$$\langle \mathrm{d}W \rangle = -\langle \mathrm{d}Q_y \rangle - \langle \mathrm{d}W_y^{\mathrm{int}} \rangle, \tag{19}$$

and

$$\langle \mathrm{d}Q_\phi(\mathbf{x}) \rangle = -\langle \mathrm{d}W_\phi^{\mathrm{int}}(\mathbf{x}) \rangle. \tag{20}$$

Further, from Eq. (7), we find

$$-\langle \mathrm{d}W_y^{\mathrm{int}} \rangle = \int \mathrm{d}^d\mathbf{x} \, \langle \mathrm{d}W_\phi^{\mathrm{int}}(\mathbf{x}) \rangle. \tag{21}$$

Then, by using Eqs. (19) and (20), this equation implies that the total dissipation of the entire process is given by the work applied to the probe, i.e.,

$$\langle \mathrm{d}W \rangle = -\int \mathrm{d}^d\mathbf{x} \, \langle \mathrm{d}Q_\phi(\mathbf{x}) \rangle - \langle \mathrm{d}Q_y \rangle, \tag{22}$$

indicating the physical consistency of our approach.

## II. STOCHASTIC ENTROPY PRODUCTION

In accordance with the framework of stochastic thermodynamics for a finite number of degrees of freedom coupled to a heat bath, and with the framework of irreversibility for fields [2, 3], we define the total entropy production along a trajectory $\{\mathbf{Y}(t), \phi(\mathbf{x}, t)\}_{t=t_\mathrm{i}}^{t_\mathrm{f}}$ of the system from time $t_\mathrm{i}$ to $t_\mathrm{f}$ as

$$S_{\mathrm{tot}}[\{\mathbf{Y}, \phi\}_{t_\mathrm{i}}^{t_\mathrm{f}}] = \ln \frac{\mathcal{P}[\{\mathbf{Y}, \phi\}_{t_\mathrm{i}}^{t_\mathrm{f}}]}{\mathcal{P}^{\mathrm{R}}[\{\mathbf{Y}^{\mathrm{R}}, \phi^{\mathrm{R}}\}_{t_\mathrm{i}}^{t_\mathrm{f}}]}, \tag{23}$$

where $\mathcal{P}$ denotes the path probability of the forward path of the joint trajectory of field and probe, starting from the probability density field $\rho_{\mathbf{Y},\phi}[\mathbf{Y}(t_\mathrm{i}), \phi(\mathbf{x}, t_\mathrm{i})]$. $\mathcal{P}^{\mathrm{R}}$ denotes the probability of the corresponding backward path in the time-reversed process, initialized with $\rho_{\mathbf{Y},\phi}[\mathbf{Y}(t_\mathrm{f}), \phi(\mathbf{x}, t_\mathrm{f})]$. In the time-reversed process, the time is running in the reverse order from $t_\mathrm{f}$ to $t_\mathrm{i}$, hence the time-dependent external driving protocols $\mathcal{U}(\mathbf{x}, t) + \mathbf{F}_{\mathrm{ext}}(t)$ are reversed in time. (In the example of a particle dragged by a harmonic trap discussed in the Main Text, this corresponds to reversing the dragging velocity from $\mathbf{v}$ to $-\mathbf{v}$.) Due to the Gaussian statistics of the noise, the path weights are proportional to the exponential of the Onsager–Machlup action $\mathcal{A}$ [2, 4], i.e.,

$$\mathcal{P} \propto \rho_{\mathbf{Y},\phi}[\mathbf{Y}(t_\mathrm{i}), \phi(\mathbf{x}, t_\mathrm{i})] e^{-\mathcal{A}[\phi, \mathbf{Y}]}. \tag{24}$$

Since the noise terms of the probe and field are assumed to be independent from each other, the total action of the joint process is the sum of the actions of probe and field, reading [2]

$$\mathcal{A} = \frac{1}{4\gamma_y T} \int_{t_\mathrm{i}}^{t_\mathrm{f}} \mathrm{d}t \, (\gamma_y \dot{\mathbf{Y}} + \boldsymbol{\nabla}_{\mathbf{Y}}\mathcal{H}^{\mathrm{int}} + \boldsymbol{\nabla}_{\mathbf{Y}}\mathcal{U} - \mathbf{F}_{\mathrm{ext}})^2 + \frac{1}{4\gamma_\phi T} \int_{t_\mathrm{i}}^{t_\mathrm{f}} \mathrm{d}t \int \mathrm{d}^d\mathbf{x} \, (\gamma_\phi \dot{\phi} + \boldsymbol{\nabla} \cdot \mathbf{J}_\mathrm{d})(-\nabla^2)^{-\mathfrak{a}} (\gamma_\phi \dot{\phi} + \boldsymbol{\nabla} \cdot \mathbf{J}_\mathrm{d}). \tag{25}$$

All the stochastic integrals in this section [such as the one appearing in Eq. (25)] are intended in the Stratonovich sense, but we will omit the $\circ$ sign to ease the notation. Accordingly, in Eq. (25) we are actually adopting the Freidlin-Wentzell discretization convention; note that the final result for the entropy production [see, c.f., Eq. (30)] is however insensitive to the choice of such convention — see, for instance, Section III in Ref. [5], or Section 5.2.2.2 in Ref. [6].

In Eq. (25) we defined

$$\boldsymbol{\nabla} \cdot \mathbf{J}_{\mathrm{d}} = (-\nabla^2)^{\mathfrak{a}} \frac{\delta \mathcal{H}}{\delta \phi}, \tag{26}$$

with $\mathfrak{a} = 0$ for non-conserved dynamics and $\mathfrak{a} = 1$ for conserved dynamics. Note that for $\mathfrak{a} = 1$ the operator $(-\nabla^2)^{\mathfrak{a}}$ is nonlocal, hence, in the last term on the r.h.s. of Eq. (25), we use $\int \mathrm{d}^d \mathbf{x}\, A(\mathbf{x})(-\nabla^2)^{-\mathfrak{a}} B(\mathbf{x})$ as a shorthand for $\int \mathrm{d}^d \mathbf{x} \int \mathrm{d}^d \mathbf{x}'\, A(\mathbf{x})(-\nabla^2)^{-\mathfrak{a}}(\mathbf{x}, \mathbf{x}') B(\mathbf{x}')$. The term $-\boldsymbol{\nabla} \cdot \mathbf{J}_{\mathrm{d}}$ in Eq. (26) is the deterministic dynamical operator of the field dynamics in Eq. (3) of the main text, i.e., $-\boldsymbol{\nabla} \cdot \mathbf{J}_{\mathrm{d}} = \gamma_\phi \dot{\phi} - \eta^{(\mathfrak{a})}$. The action of the time-reversed process is accordingly given by

$$\mathcal{A}^{\mathrm{R}} = \frac{1}{4\gamma_y T} \int_{t_{\mathrm{i}}}^{t_{\mathrm{f}}} \mathrm{d}t\, (-\gamma_y \dot{\mathbf{Y}} + \boldsymbol{\nabla}_{\mathbf{Y}}\mathcal{H}^{\mathrm{int}} + \boldsymbol{\nabla}_{\mathbf{Y}}\mathcal{U} - \mathbf{F}_{\mathrm{ext}})^2$$
$$+ \frac{1}{4\gamma_\phi T} \int_{t_{\mathrm{i}}}^{t_{\mathrm{f}}} \mathrm{d}t \int \mathrm{d}^d \mathbf{x}\, (-\gamma_\phi \dot{\phi} + \boldsymbol{\nabla} \cdot \mathbf{J}_{\mathrm{d}})(-\nabla^2)^{-\mathfrak{a}}(-\gamma_\phi \dot{\phi} + \boldsymbol{\nabla} \cdot \mathbf{J}_{\mathrm{d}}). \tag{27}$$

Inserting the path probabilities into Eq. (23) yields

$$S_{\mathrm{tot}} = \underbrace{\ln \frac{\rho_{\mathbf{Y},\phi}[\mathbf{Y}(t_{\mathrm{i}}), \phi(t_{\mathrm{i}})]}{\rho_{\mathbf{Y},\phi}[\mathbf{Y}(t_{\mathrm{f}}), \phi(t_{\mathrm{f}})]}}_{=\Delta S^{\mathrm{sh}}_{y,\phi}} + \mathcal{A}^{\mathrm{R}} - \mathcal{A}, \tag{28}$$

where the first term on the r.h.s. can be identified as the change, along the considered time interval, of the fluctuating Shannon entropy associated with the probability density field $\rho_{\mathbf{Y},\phi}$ of the joint process, i.e., considering both the probe and the field. Furthermore, the second term on the r.h.s. of Eq. (28) can be simplified to

$$\mathcal{A}^{\mathrm{R}} - \mathcal{A} = \frac{1}{\gamma_y T}\left[\int_{t_{\mathrm{i}}}^{t_{\mathrm{f}}} \mathrm{d}t\, \gamma_y \dot{\mathbf{Y}} \underbrace{(\boldsymbol{\nabla}_{\mathbf{Y}}\mathcal{H}^{\mathrm{int}} + \boldsymbol{\nabla}_{\mathbf{Y}}\mathcal{U} - \mathbf{F}_{\mathrm{ext}})}_{=(\boldsymbol{\nu} - \gamma_y \dot{\mathbf{Y}})}\right] + \frac{1}{\gamma_\phi T}\left[\int \mathrm{d}^d \mathbf{x} \int_{t_{\mathrm{i}}}^{t_{\mathrm{f}}} \mathrm{d}t\, \gamma_\phi \dot{\phi}\, (-\nabla^2)^{-\mathfrak{a}} \underbrace{\boldsymbol{\nabla} \cdot \mathbf{J}_{\mathrm{d}}}_{=(\eta^{(\mathfrak{a})} - \gamma_\phi \dot{\phi})}\right]$$
$$= \frac{1}{T}\left[\int_{\{\mathbf{Y},\phi\}_{t_{\mathrm{i}}}^{t_{\mathrm{f}}}} (\boldsymbol{\nu} - \gamma_y \dot{\mathbf{Y}})\, \mathrm{d}\mathbf{Y}\right] + \frac{1}{T}\left[\int \mathrm{d}^d \mathbf{x} \int_{\{\mathbf{Y},\phi\}_{t_{\mathrm{i}}}^{t_{\mathrm{f}}}} \mathrm{d}\phi(\mathbf{x})(-\nabla^{-2})^{\mathfrak{a}}\left(\eta^{(\mathfrak{a})} - \gamma_\phi \dot{\phi}\right)\right]. \tag{29}$$

In the last step, we have inserted the equations of motion (2) and (3) of the main text, and we converted the integrals over time into integrals along the trajectory $\{\mathbf{Y},\phi\}_{t_{\mathrm{i}}}^{t_{\mathrm{f}}}$.

As a final step, we identify in Eq. (29) the heat flows as given in Eqs. (13) and (17), and therefore obtain

$$S_{\mathrm{tot}} = -\frac{Q_y}{T} - \frac{\int \mathrm{d}^d \mathbf{x}\, Q_\phi(\mathbf{x})}{T} + \Delta S^{\mathrm{sh}}_{y,\phi}, \tag{30}$$

consistently with the total thermodynamic entropy production, given in Eq. (11) in the main text.

From Eq. (10) of the main text [or equivalently, Eq. (23)] it automatically follows that, on ensemble average, the total entropy production is constrained by the second law $\langle S_{\mathrm{tot}}\rangle \geq 0$ [7, 8]. The fluctuations of $S_{\mathrm{tot}}$ obey, by construction, the integrated and detailed fluctuation theorems, $\langle e^{-S_{\mathrm{tot}}}\rangle = 1$ and $p[S_{\mathrm{tot}}]/p[-S_{\mathrm{tot}}] = e^{-S_{\mathrm{tot}}}$ [7, 8]. Along long trajectories with $t_{\mathrm{f}} \gg t_{\mathrm{i}}$, this further implies that the total dissipation $Q = \int \mathrm{d}\mathbf{x}\, Q_\phi(\mathbf{x}) + Q_y$ asymptotically also fulfills these relations, as $\Delta S^{\mathrm{sh}}_{y,\phi}$ — contrary to $Q_\phi$ and $Q_y$ — does not grow with time.

## III. COMPARISON WITH THE DISSIPATED POWER PREDICTED BY A GENERALIZED LANGEVIN EQUATION

An established approach [9–14] to study the thermodynamics of driven particles in complex environments is based on using generalized Langevin equations (GLEs) [15, 16]. Within this description, the effect of the (slowly relaxing) medium on the particle is described via a friction kernel and a colored noise. However, GLEs generally capture only temporal correlations of the medium, but not dynamically-varying spatial correlations. In this section, we consider the conditions under which a GLE may capture the different scaling regimes in the mean dissipation rate, which we describe in the main text (see, e.g., Fig. 3 of the main text).

In particular, consider an arbitrary linear (overdamped) GLE of the form

$$\gamma \dot{Y} = -\int_{-\infty}^{t} dt' \, \Gamma(t-t') \dot{Y}(t') - \kappa(Y - vt) + \mu, \tag{31}$$

with a certain friction kernel $\Gamma$ (which we do not need to specify further here) and a zero-mean Gaussian colored noise $\mu$. We consider a nonequilibrium steady-state of the system with $v \neq 0$ and are particularly interested in the mean position (relative to the center of the trap), from which we can access the mean power [see Eq. (13) in the main text].

By averaging the GLE (31) over the possible realizations of the noise, one finds

$$\gamma \dot{u} = -\int_{-\infty}^{t} dt' \, \Gamma(t-t') \dot{u}(t') - \kappa(u - vt), \tag{32}$$

with $u(t) \equiv \langle Y(t) \rangle$. In the steady state, the average velocity $\dot{u}(t)$ of the particle does not depend on time $t$. Accordingly, we can further simplify the previous equation as

$$\gamma \dot{u} = -G_{\mathrm{m}} \dot{u} - \kappa(u - vt), \tag{33}$$

where we have introduced the constant

$$G_{\mathrm{m}} = \int_{-\infty}^{t} dt' \, \Gamma(t-t'). \tag{34}$$

The solution of Eq. (33) is

$$u = vt + \mathcal{C}, \quad \text{with} \quad \mathcal{C} = -(\gamma + G_{\mathrm{m}})v/\kappa. \tag{35}$$

Together with Eq. (21) in the main text, this implies that, for *any* linear GLE, with arbitrary friction kernel, the correction to the average dissipated power is

$$\langle \dot{W} \rangle = -\kappa v \left[\langle Y(t) \rangle - vt + v\gamma/\kappa\right] = -\kappa v \left(\mathcal{C} + v\gamma/\kappa\right) = G_{\mathrm{m}} v^2. \tag{36}$$

This expression, being quadratic in $v$, corresponds to the first regime among those shown in Fig. 3(b) in the main text, and to the usual Stokes friction. We conclude that the other regimes discussed therein can only possibly emerge in the presence of nonlinearities in the effective evolution equation of the probe.

Now, let us consider a generalization of Eq. (31) to the case of a nonlinear friction kernel [17, 18], of the generic form:

$$\gamma \dot{Y} = -\int_{-\infty}^{t} dt' \, \Gamma(t-t') f[\dot{Y}(t')] - \kappa(Y - vt) + \mu, \tag{37}$$

with some arbitrary nonlinear function $f$ [19]. Repeating the same steps as above, we find that the steady-state solution is again of the form given in Eq. (35), with

$$\mathcal{C} = -\left[\gamma v + G_{\mathrm{m}} \langle f(\dot{Y}) \rangle\right]/\kappa. \tag{38}$$

Clearly, if we allow $f$ to be nonlinear in $\dot{Y}$ (i.e., "non-Stokesian friction"), this equation can also reproduce the different regimes we found in the friction (equivalently, mean power) described in the main text. For example, the scaling in the second regime in Fig. 3(b) in the main text formally corresponds to a situation where Eq. (37) has a contribution to the friction of the form $-\int_{-\infty}^{t} dt' \, \Gamma(t-t') [\dot{Y}(t')]^{-2}$. This is a rather unusual friction term that seems difficult to justify on heuristic or phenomenological grounds alone. In contrast, by systematically integrating out the field degrees of freedom $\phi(\mathbf{x}, t)$ from the joint field–particle dynamics [given in Eqs. (2) and (3) in the main text], we can obtain an effective description for the motion of the particle in the form of a nonlinear GLE akin to Eq. (37) [see, c.f., Eq. (91)] — as discussed in the next sections.

## IV. BROWNIAN PARTICLE AND GAUSSIAN FIELD: INDEPENDENT PROCESSES

We summarize here the well-known solutions of the non-interacting processes for $\mathbf{Z}(t)$ and $\varphi_{\mathbf{p}}(t)$ [equivalently, $\mathbf{Y}(t)$ and $\phi_{\mathbf{p}}(t)$] which are obtained by setting the coupling constant $\lambda = 0$ in Eqs. (16) and (17) of the main text. This case forms the basis for the perturbative calculation of the moment generating function of the probe particle position, which we derive in Section V.

## A. Brownian particle in a harmonic potential

The equation of motion for a Brownian particle in a harmonic potential

$$\gamma_y \dot{\mathbf{Z}} = -\kappa \mathbf{Z} + \boldsymbol{\nu}, \tag{39}$$

with

$$\langle \nu_i(t)\nu_j(t') \rangle = (2T/\gamma_y)\delta_{ij}\delta(t-t'), \tag{40}$$

is that of the Ornstein-Uhlenbeck (OU) process, which we briefly revise here for clarity and to set the notation for the following sections. Each component $Z_j$ (with $j \in \{1, \ldots, d\}$) of the particle position $\mathbf{Z}$ is a Gaussian and Markovian process. Accordingly, the propagator of the process is

$$P_{1|1}(\mathbf{Z}, t|\mathbf{Z}_0, t_0) = \frac{1}{[2\pi\sigma_0^2(t, t_0)]^{d/2}} \exp\left[-\frac{|\mathbf{Z} - \mathbf{m}(t)|^2}{2\sigma_0^2(t, t_0)}\right], \tag{41}$$

where

$$\mathbf{m}(t) = \mathbf{m}(t; \mathbf{Z}_0, t_0) \equiv \langle \mathbf{Z}(t)|\mathbf{Z}(t_0) = \mathbf{Z}_0 \rangle \tag{42}$$

is the expectation value of the particle position $\mathbf{Z}(t)$ given that the process started at $\mathbf{Z}_0$ at time $t_0$, and

$$\sigma_0^2(t, t_0) \equiv \langle Z_j^2(t)|Z_j(t_0) = (\mathbf{Z}_0)_j \rangle - m_j^2(t) \tag{43}$$

is the corresponding variance. Because of isotropy, $\sigma_0^2(t, t_0)$ is independent of the component $j \in \{1, \ldots, d\}$. For the process we are currently interested in, they read

$$\mathbf{m}(t) = \mathbf{Z}_0 e^{-\kappa(t-t_0)/\gamma_y}, \tag{44}$$

$$\sigma_0^2(t, t_0) = \sigma_0^2(t - t_0) = \frac{T}{\kappa}\left[1 - e^{-2\kappa(t-t_0)/\gamma_y}\right]. \tag{45}$$

Note that the time-translational invariance of the system implies $P(\mathbf{Z}, t|\mathbf{Z}_0, t_0) = P(\mathbf{Z}, t-t_0|\mathbf{Z}_0, 0)$. Using the Langevin equation (39) or, alternatively, Eq. (41), one can readily calculate the connected two-time correlation function

$$\begin{aligned} C(t_1, t_2) &\equiv \langle Z_j(t_1)Z_j(t_2) \rangle_c = \langle [Z_j(t_1) - m_j(t_1)][Z_j(t_2) - m_j(t_2)] \rangle \\ &= \frac{T}{\kappa}\left[e^{-\kappa|t_2-t_1|/\gamma_y} - e^{-\kappa(t_1+t_2-2t_0)/\gamma_y}\right], \end{aligned} \tag{46}$$

which, again, is independent of the value of $j$, while correlations $\langle Z_j(t_1)Z_{i \neq j}(t_2) \rangle_c$ vanish. At long times, the system attains the thermal equilibrium described by the probability distribution

$$P_1^{\text{st}}(\mathbf{z}) = \frac{1}{(2\pi T/\kappa)^{d/2}} \exp\left(-\frac{\kappa z^2}{2T}\right). \tag{47}$$

## B. Dynamics of the free field

We now consider the Langevin equation for the free field, i.e., the one obtained from Eq. (3) in the main text by setting $\mathcal{H}^{\text{int}} = 0$, and using the Gaussian $\mathcal{H}_\phi$ given in Eq. (14) In Fourier space, this equation reads

$$\gamma_\phi \dot{\phi}_{\mathbf{p}} = -\gamma_\phi \alpha_p \phi_{\mathbf{p}} + \eta_{\mathbf{p}}, \tag{48}$$

where, as in the main text, we conveniently introduced $\alpha_p = p^{2\mathfrak{a}}(p^2 + r)/\gamma_\phi$, with $\mathfrak{a} = 0$ or $1$ corresponding to non-conserved or conserved dynamics, respectively. The zero-average Gaussian noise $\eta_{\mathbf{p}}$ above is characterized by the correlations

$$\langle \eta_{\mathbf{p}}(t)\eta_{\mathbf{p}'}(t') \rangle = \gamma_\phi^2 \Omega_\phi(p)\delta^d(p+p')\delta(t-t'), \tag{49}$$

where $\Omega_\phi(p) = 2Tp^{2\mathfrak{a}}/\gamma_\phi$. We adopted the Fourier convention $f(\mathbf{x}) = \int [\mathrm{d}^d p/(2\pi)^d] e^{i\mathbf{p}\cdot\mathbf{x}} f_\mathbf{p}$, with the delta distribution normalized as $\int [\mathrm{d}^d p/(2\pi)^d] \delta^d(\mathbf{p}) = 1$. Equation (48) is the same as the one describing the position of a Ornstein-Uhlenbeck particle, whence

$$\langle \phi_{p_1}(s_1) \phi_{p_2}(s_2) \rangle = \delta^d(p_1 + p_2) C^{(0)}_{p_1}(s_1, s_2), \tag{50}$$

$$C^{(0)}_p(s_1, s_2) = \frac{T}{p^2 + r} \left[ e^{-\alpha_p |s_2 - s_1|} - e^{-\alpha_p (s_1 + s_2 - 2t_0)} \right]. \tag{51}$$

In the previous expressions, we introduced the superscript (0) in order to distinguish these expressions from those derived in, c.f., Section V, which pertain to a reference frame moving at velocity $\mathbf{v}$. Note that, by construction, $C^{(0)}_p(s_1, s_2) = C^{(0)}_p(s_2, s_1)$. In the limit in which the time $t_0$ of initialization of the field is sent to $t_0 \to -\infty$, one recovers the equilibrium correlator

$$C^{(0)}_p(\tau) = \frac{T}{p^2 + r} e^{-\alpha_p |\tau|}, \tag{52}$$

which is a function of the time difference $\tau = s_2 - s_1$ only. It is also customary to define the response function and linear susceptibility of the free field [20]

$$\chi^{(0)}_p(s_1, s_2) = \chi^{(0)}_p(s_2 - s_1) = \frac{p^{2\mathfrak{a}}}{\gamma_\phi} G^{(0)}_p(s_2 - s_1), \tag{53}$$

$$G^{(0)}_p(s_2 - s_1) = e^{-\alpha_p (s_2 - s_1)} \Theta(s_2 - s_1), \tag{54}$$

respectively, where $\Theta(t)$ is the Heaviside theta function. The functions $\chi^{(0)}_p(t)$ and $G^{(0)}_p(t)$ are time-translational invariant (also in a generic, possibly nonequilibrium, stationary state, since the equation of motion is). Moreover, in equilibrium, they are linked to the correlator in Eq. (52) by the fluctuation-dissipation theorem [20]

$$T\chi^{(0)}_p(\tau) = -\Theta(\tau) \frac{\partial}{\partial \tau} C^{(0)}_p(\tau). \tag{55}$$

## V. DERIVATION OF THE MOMENT GENERATING FUNCTION OF THE PARTICLE POSITION

Here we provide the derivation of the cumulant generating function of the particle position reported in Eq. (18) of the main text. We start from the coupled equations of motion for the variables $\mathbf{Z} = \mathbf{Y} - \mathbf{v}t + \mathbf{v}\gamma_y/\kappa$ and $\varphi(\mathbf{z}, t) \equiv \phi(\mathbf{z} + \mathbf{v}t - \mathbf{v}\gamma_y/\kappa, t)$, i.e.,

$$\gamma_y \dot{\mathbf{Z}}(t) = -\kappa \mathbf{Z} + \lambda \int \frac{\mathrm{d}^d p}{(2\pi)^d} i\mathbf{p} V_{-p} \varphi_\mathbf{p} e^{i\mathbf{p}\cdot\mathbf{Z}} + \boldsymbol{\nu}(t), \tag{56}$$

$$\gamma_\phi (\partial_t - i\mathbf{p}\cdot\mathbf{v} + \alpha_p) \varphi_\mathbf{p} = \lambda p^{2\mathfrak{a}} V_p e^{-i\mathbf{p}\cdot\mathbf{Z}} + \eta_\mathbf{p}. \tag{57}$$

Note that, for $\lambda = 0$, the second equation is formally the same as Eq. (48) with $\alpha_p \mapsto \alpha_p - i\mathbf{p}\cdot\mathbf{v}$, which describes the free Gaussian field. Its solution in the stationary state thus remains the same upon replacing the equilibrium correlator $C^{(0)}_p(t)$ and the free susceptibility $\chi^{(0)}_p(t)$ in Eqs. (52) and (53) by

$$C_\mathbf{p}(t) = \frac{T}{p^2 + r} e^{-(\alpha_p - i\mathbf{p}\cdot\mathbf{v})|t|} = e^{i\mathbf{p}\cdot\mathbf{v}|t|} C^{(0)}_p(t), \tag{58}$$

$$\chi_\mathbf{p}(t) = \frac{p^{2\mathfrak{a}}}{\gamma_\phi} G_\mathbf{p}(t), \tag{59}$$

$$G_\mathbf{p}(t) = e^{-(\alpha_p - i\mathbf{p}\cdot\mathbf{v})t} \Theta(t) = e^{i\mathbf{p}\cdot\mathbf{v}t} G^{(0)}_p(t). \tag{60}$$

Starting from the Langevin equations (56) and (57), and assuming the coupling constant $\lambda$ to be small, it is possible to derive a (forward) evolution equation which describes the marginal probability distribution $P_1(\mathbf{z}, t)$ of the particle coordinate alone. The derivation is presented in Ref. [21], where a more general case is treated, in which a second particle also interacts with the field via an interaction potential $V^{(z)}_p$; the present case corresponds to $V^{(z)}_p \mapsto 0$. We report here only the final result:

$$\partial_t P_1(\mathbf{z}, t) = \mathcal{L}_0(\mathbf{z}) P_1(\mathbf{z}, t) + \lambda^2 \int_{-\infty}^t \mathrm{d}s \int \mathrm{d}\mathbf{x}\, \mathcal{L}(\mathbf{z}, \mathbf{x}; t - s) P_2(\mathbf{z}, t; \mathbf{x}, s) + \mathcal{O}(\lambda^4). \tag{61}$$

Here
$$\mathcal{L}_0 \equiv \boldsymbol{\nabla}_{\mathbf{z}} \cdot \left( \frac{\kappa}{\gamma_y} \mathbf{z} + \frac{T}{\gamma_y} \boldsymbol{\nabla}_{\mathbf{z}} \right) \tag{62}$$

is the Fokker-Planck operator for an Ornstein-Uhlenbeck particle [22], while the second term on the r.h.s. of Eq. (61) involves a convolution of the two-time probability distribution $P_2(\mathbf{z}, t; \mathbf{x}, s)$ with a memory kernel $\mathcal{L}(\mathbf{r}, u)$. This typically emerges in non-Markovian processes, where one usually obtains a hierarchy of master equations relating the $n$-point distribution $P_n(\mathbf{x_n}, t_n; \mathbf{x_{n-1}}, t_{n-1}; \ldots; \mathbf{x_1}, t_1)$ with $P_{n+1}$ (see for instance Refs. [23–26]). This kernel reads

$$\mathcal{L}(\mathbf{z}, \mathbf{x}; u) \equiv \gamma_y^{-1} \nabla_{\mathbf{z}}^l \int \frac{\mathrm{d}^d p}{(2\pi)^d} i p_l |V_p|^2 e^{-i\mathbf{p} \cdot (\mathbf{z} - \mathbf{x})} \left[ \chi_{\mathbf{p}}(u) - \frac{i}{\gamma_y} C_{\mathbf{p}}(u) e^{-\kappa u / \gamma_y} p_j \nabla_{\mathbf{z}}^j \right], \tag{63}$$

where the notation $\nabla_{\mathbf{z}}^j$ is shorthand for $\partial/\partial z_j$, and the summation over the repeated indices $j$ and $l$ is implied. This last expression had already been reported in Ref. [21] but, unfortunately, with a confusing notation.

We want to find an expression for the characteristic function

$$g(\mathbf{k}, t) \equiv \langle e^{-i\mathbf{k} \cdot \mathbf{Z}(t)} \rangle = \int \mathrm{d}\mathbf{z} \, e^{-i\mathbf{k} \cdot \mathbf{z}} P_1(\mathbf{z}, t). \tag{64}$$

We start by taking the Fourier transform of the various terms in the master equation (61), which leads to

$$\left[ \partial_t - \mathcal{L}_0^\dagger(\mathbf{k}) \right] g(\mathbf{k}, t) = f(\mathbf{k}, t) + \mathcal{O}(\lambda^4), \tag{65}$$

where, from Eq. (62),

$$\mathcal{L}_0^\dagger(\mathbf{k}) = -\frac{\kappa}{\gamma_y} \mathbf{k} \cdot \boldsymbol{\nabla}_{\mathbf{k}} - \frac{T}{\gamma_y} k^2, \tag{66}$$

and we have introduced

$$f(\mathbf{k}, t) \equiv \lambda^2 \int_{-\infty}^t \mathrm{d}s \int \mathrm{d}\mathbf{x} \, \mathrm{d}\mathbf{z} \, e^{-i\mathbf{k} \cdot \mathbf{z}} \mathcal{L}(\mathbf{z}, \mathbf{x}; t - s) P_2(\mathbf{z}, t; \mathbf{x}, s). \tag{67}$$

The solution of Eq. (65) is formally given, at the lowest nontrivial order in $\lambda$, by

$$g(\mathbf{k}, t) = g^{(0)}(\mathbf{k}) + \int_{-\infty}^t \mathrm{d}t' \int \frac{\mathrm{d}^d k'}{(2\pi)^d} g_{1|1}(\mathbf{k}, t|\mathbf{k}', t') f(\mathbf{k}', t') + \mathcal{O}(\lambda^4), \tag{68}$$

where we introduced

$$g^{(0)}(\mathbf{p}) \equiv \exp(-p^2 T / 2\kappa), \tag{69}$$

i.e., the moment generating function of the uncoupled ($\lambda = 0$) particle, and where we indicated by $g_{1|1}$ the propagator of the Ornstein-Uhlenbeck process in momentum space. The latter is easily found by solving the homogeneous part of Eq. (65) via the method of characteristics [27], or even by simply Fourier transforming the real space propagator given in Eq. (41): both ways lead to

$$g_{1|1}(\mathbf{k}, t|\mathbf{k}', t') = e^{-(k^2/2)\sigma_0^2(t-t')} \delta^d(\mathbf{k}' + \mathbf{k} e^{-\kappa(t-t')/\gamma_y}), \tag{70}$$

where $\sigma_0(t - t')$ was introduced in Eq. (45).

Let us now focus on the stationary state, for which

$$g(\mathbf{k}, t) \to g(\mathbf{k}), \qquad g_{1|1}(\mathbf{k}, t|\mathbf{k}', t') \to g_{1|1}(\mathbf{k}, t - t'|\mathbf{k}', 0), \qquad \text{and} \qquad P_2(\mathbf{z}, t; \mathbf{x}, s) \to P_2(\mathbf{z}, t - s; \mathbf{x}, 0). \tag{71}$$

Equation (68) then becomes

$$\begin{aligned} g(\mathbf{k}) &= g^{(0)}(\mathbf{k}) + \int_0^\infty \mathrm{d}u' \int \frac{\mathrm{d}^d k'}{(2\pi)^d} g_{1|1}(\mathbf{k}, u'|\mathbf{k}', 0) f(\mathbf{k}') + \mathcal{O}(\lambda^4) \\ &= g^{(0)}(\mathbf{k}) + \int_0^\infty \mathrm{d}u' \, e^{-(k^2/2)\sigma_0^2(u')} f\left(-\mathbf{k} e^{-\kappa u'/\gamma_y}\right) + \mathcal{O}(\lambda^4), \end{aligned} \tag{72}$$

where in the second line we used Eq. (70). Similarly, manipulating Eqs. (63) and (67) renders

$$f(\mathbf{k}) = -\frac{\lambda^2}{\gamma_y}\int_0^\infty \mathrm{d}u \int \frac{\mathrm{d}^d p}{(2\pi)^d}|V_p|^2(\mathbf{p}\cdot\mathbf{k})\left[\chi_{\mathbf{p}}(u) + \frac{\mathbf{p}\cdot(\mathbf{p}+\mathbf{k})}{\gamma_y}C_{\mathbf{p}}(u)e^{-\kappa u/\gamma_y}\right]\widetilde{P}_2(\mathbf{p}+\mathbf{k},u;-\mathbf{p},0) \tag{73}$$

in terms of the double Fourier transform $\widetilde{P}_2$ of the two-point probability distribution $P_2$, with respect to both its spatial arguments. Since we are neglecting terms of order higher than $\lambda^2$, then $P_2$ above can be replaced by its expression for $\lambda = 0$, i.e.,

$$P_2(\mathbf{z},t;\mathbf{x},s) = P_{1|1}(\mathbf{z},t|\mathbf{x},s)P_1(\mathbf{x},s) + \mathcal{O}(\lambda^2), \tag{74}$$

where $P_{1|1}$ is the Ornstein-Uhlenbeck propagator given in Eq. (41), with Fourier transform $g_{1|1}$ provided in Eq. (70). In the long-time limit, we can replace $P_1(\mathbf{x},s)$ by the stationary distribution $P_1^{\mathrm{st}}(\mathbf{x})$ eventually attained by the uncoupled particle in its harmonic trap, see Eq. (47). We recall that the Fourier transform of $P_1^{\mathrm{st}}(\mathbf{x})$ is the function $g^{(0)}(\mathbf{p})$ given in Eq. (69). This yields

$$\widetilde{P}_2(\mathbf{p}_2,t;\mathbf{p}_1,0) \simeq \int \frac{\mathrm{d}^d p_0}{(2\pi)^d} g_{1|1}(\mathbf{p}_2,t|\mathbf{p}_1-\mathbf{p}_0,0)g^{(0)}(\mathbf{p}_0) = g^{(0)}(\mathbf{p}_1+\mathbf{p}_2 e^{-\kappa t/\gamma_y})\exp\left[-\frac{p_2^2}{2}\sigma_0^2(t)\right]. \tag{75}$$

Equation (73) can be further simplified by noting that, for any $u$-independent value $A$ [and, in particular, for $A \equiv \mathbf{p}\cdot(\mathbf{p}+\mathbf{k})$], we can rewrite [28]

$$\left[\chi_p^{(0)}(u) + \frac{A}{\gamma_y}C_p^{(0)}(u)e^{-\kappa u/\gamma_y}\right] = -\frac{e^{A\sigma^2(u)}}{\alpha_p}\frac{\partial}{\partial u}\left[\chi_p^{(0)}(u)e^{-A\sigma^2(u)}\right], \tag{76}$$

where we recall that $\chi_p^{(0)}(u)$ and $C_p^{(0)}(u)$ differ from the corresponding $\chi_{\mathbf{p}}(u)$ and $C_{\mathbf{p}}(u)$ only by a multiplicative factor $\exp(i\mathbf{p}\cdot\mathbf{v}\,u)$ for $u > 0$ [see Eqs. (58) to (60)], and where we introduced

$$\sigma^2(u) \equiv \sigma_0^2(u/2), \tag{77}$$

with $\sigma_0$ given in Eq. (45). Equation (76) readily follows as a direct consequence of the fluctuation-dissipation theorem for the free field, recalled in Eq. (55). From Eq. (73), after integrating by parts in $u$, we thus get

$$f(\mathbf{k}) = -i\frac{\lambda^2}{\gamma_y}g^{(0)}(\mathbf{k})\int_0^\infty \mathrm{d}u \int \frac{\mathrm{d}^d p}{(2\pi)^d}\frac{|V_p|^2(\mathbf{p}\cdot\mathbf{k})(\mathbf{p}\cdot\mathbf{v})}{p^2+r}G_{\mathbf{p}}(u)e^{-\mathbf{p}\cdot(\mathbf{p}+\mathbf{k})\sigma^2(u)}, \tag{78}$$

where $G_{\mathbf{p}}(u)$ is the field response propagator in the comoving frame, see Eq. (60). We finally substitute the expression of $f(\mathbf{k})$ given in Eq. (78) into Eq. (72) for $g(\mathbf{k})$, and we simplify the integral over $u'$ by recognizing that

$$(\mathbf{p}\cdot\mathbf{k})e^{-\kappa u'/\gamma_y}\exp\left[(\mathbf{p}\cdot\mathbf{k})\sigma^2(u)e^{-\kappa u'/\gamma_y}\right] = -\frac{\gamma_y}{\kappa\sigma^2(u)}\frac{\partial}{\partial u'}\exp\left[(\mathbf{p}\cdot\mathbf{k})\sigma^2(u)e^{-\kappa u'/\gamma_y}\right]. \tag{79}$$

This gives the moment generating function

$$g(\mathbf{k}) = g^{(0)}(\mathbf{k})\left\{1 + \frac{i\lambda^2}{\kappa}\int_0^\infty \frac{\mathrm{d}u}{\sigma^2(u)}\int \frac{\mathrm{d}^d p}{(2\pi)^d}\frac{(\mathbf{p}\cdot\mathbf{v})}{p^2+r}|V_p|^2 G_{\mathbf{p}}(u)e^{-p^2\sigma^2(u)}\left[1-e^{-(\mathbf{p}\cdot\mathbf{k})\sigma^2(u)}\right]\right\} + \mathcal{O}(\lambda^4). \tag{80}$$

Upon taking the logarithm of Eq. (80) and expanding for small $\lambda$, we obtain the cumulant generating function reported in Eq. (18) of the main text. In particular, computing $-\nabla_{\mathbf{q}}^2 g(\mathbf{q})|_{\mathbf{q}=\mathbf{0}}$ gives the correlation function

$$\langle Z_l Z_m\rangle_c = \frac{T}{\kappa}\delta_{lm} + \frac{\lambda^2}{\kappa}\int \frac{\mathrm{d}^d p}{(2\pi)^d}\frac{i p_l p_m(\mathbf{p}\cdot\mathbf{v})}{p^2+r}|V_p|^2\int_0^\infty \mathrm{d}u\, G_{\mathbf{p}}(u)\sigma^2(u)e^{-p^2\sigma^2(u)}. \tag{81}$$

We note that the contribution $\propto \lambda^2$ actually vanishes for $l \neq m$, as it follows from the fact that, otherwise, the integrand would be an odd function of some component of the vector $\mathbf{p}$. It thus appears that no cross-correlations arise between the various orthogonal components of the displacement, i.e., $\langle Z_l Z_m\rangle_c \propto \delta_{lm}$, but the variance is anisotropically modified along the directions parallel and perpendicular to the trap velocity $\mathbf{v}$ (see also Fig. 1). In the critical case, as well as for small velocities or small thermal fluctuations, it is possible to prove analytically that the correction in Eq. (81) is positive. Note that in the absence of the field (i.e., for $\lambda = 0$) the variance *does not* change, at finite $v$, compared to the case $v = 0$.

## VI. AVERAGE PARTICLE POSITION AND DISSIPATION RATE WITH NON-CONSERVED DYNAMICS

In this Appendix, we analyze the behavior of the average particle position $\langle \mathbf{Z} \rangle$ reported in Eq. (19) in the main text, together with its consequences on the dissipation rate discussed therein, focusing on the case of non-conserved dynamics.

### A. Typical time and length scales

First, it is useful to introduce the following timescales:

$$\tau_\kappa \equiv \gamma_y/\kappa, \quad \tau_v \equiv R/v, \quad \tau_R \equiv \gamma_\phi R^z, \quad \tau_\xi = \gamma_\phi \xi^z, \tag{82}$$

where $z = 2 + 2\mathfrak{a}$ is the dynamical critical exponent of the field ($z = 2$ for non-conserved dynamics) and $\xi$ the correlation length of its fluctuations. We recall that $R$ is the length scale which characterizes the interaction potential and thus sets the particle "radius" [see Eq. (15) in the main text], $\kappa$ is the stiffness of the harmonic trap, while the parameters $\gamma_{y,\phi}$ are the friction coefficients of the particle and the field, respectively [see Eqs. (2) and (3) in the main text].

The first timescale $\tau_\kappa$ is simply the relaxation time of the particle in its harmonic trap. The second is the time $\tau_v$ taken by the moving trap to cover a distance corresponding to the particle radius; equivalently, $\tau_v^{-1}$ estimates the shear rate near the driven particle [29]. The next two scales, instead, characterize the field $\phi$. In particular, $\tau_R$ is the time that a critical field $\phi$ takes to relax across a distance $\sim R$; similarly, $\tau_\xi$ is the relaxation time of the field $\phi$ across its correlation length $\xi$, and it represents its slowest timescale in the vicinity of the critical point (for non-conserved dynamics).

Upon appropriately rescaling the integration variables, both the cumulant generating function in Eq. (18) of the main text and the moments of the particle position $\mathbf{Z}$ can be recast in the form of scaling functions which depend only on dimensionless ratios of the typical time and length scales of the system. Consider, for instance, the average position in Eq. (19) of the main text in the case of non-conserved dynamics, where the integral over momenta $\mathbf{p}$ can be computed upon choosing a Gaussian interaction potential $V_p = \exp(-p^2 R^2/2)$. One possible choice is to rescale momenta by $\xi^{-1}$, so that after some manipulation one gets

$$\langle \mathbf{Z} \rangle = (\mathbf{v}\tau_\xi) \times \frac{\lambda^2}{\kappa \xi^d} \times g\left(\chi, \frac{R}{\xi}, \frac{\sqrt{T/\kappa}}{\xi}, \frac{\tau_\xi}{\tau_\kappa}\right), \tag{83}$$

where the second factor highlighted on the r.h.s. is dimensionless. Similarly, we introduced above the auxiliary function

$$g(\chi, \pi_1, \pi_2, \pi_3) \equiv \frac{1}{4(4\pi)^{d/2}} \int_0^\infty \mathrm{d}a \, \mathrm{d}b \, \frac{b^2\chi - 2f(a,b,\{\pi_i\})}{[f(a,b,\{\pi_i\})]^{2+d/2}} \exp\left[-a - b - \frac{b^2\chi}{4f(a,b,\{\pi_i\})}\right], \tag{84}$$

$$f(a,b,\{\pi_i\}) \equiv a + b + \pi_1^2 + \pi_2^2 \left(1 - e^{-b\pi_3}\right), \tag{85}$$

which depends on dimensionless parameters, while the parameter

$$\chi \equiv \frac{\tau_R \tau_\xi}{\tau_v^2} = (\gamma_\phi v \xi)^2 \tag{86}$$

can be interpreted as the squared Péclet number of the system, i.e., $\chi = (\text{Pe})^2$. Alternatively, we note that in the context of microrheology experiments conducted in viscoelastic media [29] one can identify the Weissenberg number $\text{Wi} \equiv \tau_s/(2\tau_v)$, where $\tau_s$ is the (single) relaxation timescale of the medium. It follows that the parameter $\chi$ in Eq. (86) can be seen as the product of two Weissenberg numbers, one for each of the two dominant relaxation timescales of the field, i.e., $\tau_R$ and $\tau_\xi$. The remaining dimensionless parameters $\{\pi_i\}$ encode the ratios of the typical length or time scales of the system. In the limit in which $\tau_\xi \gg \tau_\kappa$ (strong confinement [30, 31], or close to criticality), the function $f(a,b,\{\pi_i\})$ in Eq. (85) simplifies, and the dependence on the three parameters $\{\pi_i\}$ can be replaced by that on a single parameter $\pi_4$ defined as

$$\pi_4 \equiv \ell_T/\xi, \tag{87}$$
$$\ell_T^2 \equiv R^2 + T/\kappa. \tag{88}$$

Note that $\ell_T$ may be regarded as the effective particle radius, once its average thermal fluctuations have been taken into account: indeed, $\sqrt{T/\kappa}$ is the typical thermal fluctuation of the particle position in a trap at thermal equilibrium. We thus obtain in this limit

$$\langle \mathbf{Z} \rangle = \mathbf{v}\tau_\xi \frac{\lambda^2}{\kappa \xi^d} g'(\chi, \ell_T/\xi), \tag{89}$$

$$f'(a, b, \pi_4) = a + b + \pi_4^2, \tag{90}$$

with the function $g'$ defined as $g$ in Eq. (84), but with $f$ replaced by $f'$ above.

### B. Behavior as a function of the drag speed $v$

The average position $\langle Z \rangle$ in the steady state is plotted in the inset of Fig. 3(b) in the main text as a function of the trap velocity $v$, for the case of non-conserved dynamics in $d = 1$. For any finite value of $r$, i.e., of the correlation length $\xi = r^{-1/2}$, one can generically identify three regimes and a crossover:

- *Low-velocity (adiabatic) regime*, for $0 < v \lesssim v_I$. When $v = 0$ the system is in equilibrium, and the average particle position is not modified by the presence of the field [28], thus yielding $\langle Z(v = 0) \rangle = 0$. Inspecting Eq. (19) of the main text shows that $\langle Z \rangle \propto -v$, i.e., a linear dependence of $\langle Z \rangle$ on $v$ for small $v$.

- *Intermediate crossover*, for $v_I \lesssim v \lesssim v_{II}$. Here $|\langle Z \rangle|$ reaches a maximum and behaves in a possibly non-monotonic way (depending on the choice of the various parameters).

- *High-velocity (adiabatic) regime*, for $v_{II} \lesssim v \lesssim v_{III}$. The amplitude of the average position starts decaying as $|\langle Z \rangle| \propto v^{-1}$.

- *"Depinning" regime*, for $v \gtrsim v_{III}$. At very high speeds, one can check that $\langle Z \rangle$ tends to a small but finite value.

These regimes correspond to those observed for the correction $\propto \lambda^2$ to the dissipation rate $\langle \dot{W} \rangle_\lambda \propto v \langle Z \rangle$ in Fig. 3(b) in the main text, which grows like $\propto v^2$ for small $v$ (as it is the case for the usual Stokes friction), while it is $v$-independent for intermediate $v$, and finally grows like $\propto v$ in the last regime.

We can understand these behaviors starting from the effective equation of motion for the particle coordinate $\mathbf{Z}(t)$, which can be obtained from Eqs. (56) and (57) by solving for $\varphi_\mathbf{p}(t)$ in the second (linear) equation, and by inserting this result into the first equation [21, 30, 32]. This yields

$$\gamma_y \dot{\mathbf{Z}}(t) = -\kappa \mathbf{Z}(t) + \boldsymbol{\nu}(t) + \lambda \int \frac{\mathrm{d}^d p}{(2\pi)^d} i\mathbf{p} V_{-p} e^{i\mathbf{p}\cdot\mathbf{Z}(t)} \left[ \zeta_\mathbf{p}(t) + \lambda V_p \int_{-\infty}^t \mathrm{d}s\, \chi_\mathbf{p}(t-s) e^{-i\mathbf{p}\cdot\mathbf{Z}(s)} \right], \tag{91}$$

where the field susceptibility $\chi_\mathbf{p}(u)$ is given in Eq. (59), while we introduced the colored Gaussian noise

$$\zeta_\mathbf{p}(t) \equiv \int_{-\infty}^t \mathrm{d}s\, G_\mathbf{p}(t-s) \eta_\mathbf{p}(s), \tag{92}$$

which has zero mean and the correlator $C_\mathbf{p}(t)$ [see Eq. (58)]. Equation (91) is markedly non-Markovian because of the presence of a (nonlinear) memory term [i.e., the term that is proportional to $\chi_\mathbf{p}(u)$], and a colored multiplicative noise term $\zeta_\mathbf{p}(t)$. In the limits of very small and very large drag speed $v$, however, we physically expect to recover an approximately Markovian description. First, for very small $v$, the field is fast enough to equilibrate with the particle being fixed at its actual position, so that an adiabatic approximation is applicable [21, 28, 33, 34]. Second, for very large $v$, the evolution of the field is so slow that the particle effectively encounters a field landscape that is effectively static, i.e., quenched.

To make these statements more substantial, it is interesting to investigate the properties of the nonequilibrium steady state of the field, characterized by the formation of a comoving stationary profile around the particle, which we dubbed *shadow* in the main text. Its analytical expression is found by averaging Eq. (17) of the main text over thermal fluctuations and by setting $\partial_t \langle \varphi_\mathbf{p} \rangle = 0$, which yields, at the leading order in $\lambda$ [31],

$$\langle \varphi_\mathbf{p} \rangle_{\mathrm{st}} = \frac{\lambda}{\gamma_\phi} \frac{p^{2\mathfrak{a}} V_p \exp\left[-Tp^2/(2\kappa)\right]}{\alpha_p - i\mathbf{p}\cdot\mathbf{v}}, \tag{93}$$

where we used Eq. (18) of the main text, for $\lambda = 0$, in order to express the average $\langle \exp(-i\mathbf{p}\cdot\mathbf{Z}) \rangle = g(\mathbf{p})$ at the numerator of the resulting expression. The shadow is plotted in Fig. 3(a) of the main text for the case of $d = 1$

and non-conserved dynamics. Qualitatively, as the value of the speed $v$ increases, the shadow becomes increasingly asymmetric (with its wavefront becoming steeper and its wake longer), and its overall amplitude decreases towards zero. In the limit of a point-like particle — which is obtained by choosing an interaction potential $V(\mathbf{x}) = \delta(\mathbf{x})$ — one can explicitly compute the inverse Fourier transform of Eq. (93) in $d = 1$ non-conserved dynamics by complex integration, finding

$$\langle \varphi(z) \rangle_{\text{st}} = \frac{\lambda}{\sqrt{(v\gamma_\phi)^2 + (2/\xi)^2}} \exp\left\{ -\frac{v\gamma_\phi}{2} \left[ z + |z|\sqrt{1 + \left(\frac{2}{v\xi\gamma_\phi}\right)^2} \right] \right\}. \tag{94}$$

Accordingly, the shadow decays exponentially, in front or behind the particle (i.e., for $z \to \pm\infty$), over two generically distinct length scales

$$\ell_\pm = \xi \left[ \sqrt{1 + (\xi v \gamma_\phi/2)^2} \mp (\xi v \gamma_\phi/2) \right], \tag{95}$$

respectively, as reported in the main text. From this expression of $\ell_\pm$ it appears that $1/(\gamma_\phi \xi)$ is a natural velocity scale of this problem and, in fact, we shall see below that it coincides with the velocity $v_I$ under which the dynamics is adiabatic. Close to equilibrium, i.e., for $v \ll v_I = 1/(\gamma_\phi \xi)$, Eq. (94) reduces to

$$\langle \varphi(z) \rangle_{\text{st}} \simeq \frac{\lambda \xi}{2} e^{-|z|/\xi}, \tag{96}$$

which is symmetric as expected, and extends over distances of $\mathcal{O}(\xi)$. Conversely, for large speeds $v \gg v_I$ the amplitude of the shadow decreases as $1/v$, and decays in its front over a typical length $1/(v\gamma_\phi) \ll \xi$, and in its wake over the length $\xi^2 v \gamma_\phi \gg \xi$. Notably, by using a more realistic interaction potential $V(\mathbf{x})$ that takes into account the finite particle radius $R \ll \xi$, Eq. (94) would still describe the tails of the shadow, i.e., its behavior for $|z| \gg R$.

As we emphasized in the main text, choosing positive values of the coupling constant $\lambda$ and $V(\mathbf{x})$ in Eq. (15) results in an effective attraction between the particle and the shadow; we are interested here in how this attraction modifies the steady-state average particle position $\langle \mathbf{Z} \rangle$. Let us assume initially that this *deterministic* attraction dominates over the (non-Gaussian) thermal fluctuations induced by the field, which are encoded in the noise term $\zeta_\mathbf{p}(t)$ in Eq. (92). Focusing on the two Markovian limits described above, we can approximately replace the field $\varphi_\mathbf{p}$ in Eq. (56) by its mean value, i.e., by the shadow given in Eq. (93). This approximation implies that

$$\gamma_y \partial_t \langle \mathbf{Z}(t) \rangle \simeq -\kappa \langle \mathbf{Z} \rangle + \lambda \int \frac{\mathrm{d}^d p}{(2\pi)^d} i\mathbf{p} \left\langle e^{i\mathbf{p} \cdot \mathbf{Z}} \right\rangle \langle \varphi_\mathbf{p} \rangle_{\text{st}}. \tag{97}$$

The second term on the right hand side represents a field-induced effective force which, in the steady state defined by $\partial_t \langle \mathbf{Z}(t) \rangle = 0$, counterbalances the restoring force $-\kappa \langle \mathbf{Z} \rangle$ of the harmonic trap. For small speeds $v$, the shadow is essentially symmetric [see Eq. (96)] around its center which we will denote as $z_p$. Linearizing Eq. (97) around the unperturbed position $\langle Z \rangle = 0$ (for simplicity, we consider below the case $d = 1$) renders in general

$$\gamma_y \partial_t \langle Z \rangle \simeq -\kappa \langle Z \rangle - \lambda^2 \kappa_\phi \left( \langle Z \rangle - z_p \right), \tag{98}$$

where the expression of the effective linear strength $\kappa_\phi$ and of $z_p$ introduced above can be found on the basis of the expression of the shadow in Eq. (93) — as they are not necessary for our argument, we do not provide them here. In the steady state and at leading order in the coupling $\lambda$, we thus find

$$\langle Z \rangle \simeq \lambda^2 \frac{\kappa_\phi}{\kappa} z_p. \tag{99}$$

At small but nonzero $v$, one expects the center of the shadow to lag behind the average particle position (which effectively "drags" the shadow), i.e., $z_p = av + \mathcal{O}(v^2)$ with some $a < 0$. Accordingly, Eq. (99) with this expression for $z_p$ predicts that $|\langle Z \rangle|$ has a linear dependence on $v$ for small velocities, as observed in Fig. 3(b) in the main text (see the inset), and as discussed above.

The simplified description presented above is expected to become less accurate as $v$ increases, because the rearrangement of the field can no longer be assumed to be instantaneous, and the particle actually moves considerably (due to the dragging) while the shadow forms. The spatial extension of the shadow is of $\mathcal{O}(\xi)$ in this regime [see Eq. (96)], so it builds over a timescale $\tau_\xi \simeq \gamma_\phi \xi^2$ in the case of non-conserved dynamics [see Eq. (82)]. This timescale should be compared with the time $\tau$ taken by the trap to span a distance of the order of the field correlation length $\xi$, i.e., $\tau \simeq \xi/v$. The two timescales balance at

$$v_I \simeq 1/(\gamma_\phi \xi), \tag{100}$$

which we identify as the end of the first regime, i.e., the beginning of the crossover in Fig. 3(b) of the main text. This estimate assumes also that $\tau_\xi \ll \tau_\kappa$, i.e., that the field relaxes faster than the typical timescale set by the harmonic trap (which is typically the case in experimental realizations with colloidal particles — see Section VII B).

For intermediate velocities $v_I \leq v \leq v_{II}$, the behavior of $\langle Z \rangle$ is fully determined by non-Markovian effects which, contrary to the previous regimes, cannot be explained in terms of a simplified model. Conversely, for $v > v_{II}$, we can again replace the field by its expectation value as in Eq. (97). In this high-velocity regime, the shadow is so asymmetric [see Fig. 3(a) of the main text] that it is no longer meaningful to treat it as a point-particle concentrated in its center $z_p$ [as we essentially did in Eq. (98)]. In spite of this difficulty, it actually turns out that its form can still be determined analytically. In particular, from Eq. (93) one can first compute

$$\partial_x \langle \varphi(x) \rangle_{\text{st}} = \partial_x \int \frac{\mathrm{d}p}{2\pi} \langle \varphi_p \rangle_{\text{st}} e^{ipx} = \frac{\lambda}{\gamma_\phi} \int \frac{\mathrm{d}p}{2\pi} \frac{ip V_p e^{-Tp^2/(2\kappa)}}{\alpha_p - ipv} e^{ipx} \simeq -\frac{\lambda}{\gamma_\phi v} \int \frac{\mathrm{d}p}{2\pi} V_p e^{-Tp^2/(2\kappa)} e^{ipx}, \tag{101}$$

where the approximation introduced in the last equality is expected to be valid when $v$ exceeds *all* the relevant ratios of length and time scales of the system. This requires, inter alia, that

$$\tau_v \ll \tau_\kappa \tag{102}$$

[see Eq. (82)], which justifies the replacement of the field $\varphi_\mathbf{p}$ by its mean value in Eq. (97). Moreover, we assume (as done above and in the main text) that the interaction potential $V(x)$ is characterized by a single length scale $R$ (i.e., the particle radius), so that $V_p$ provides an effective cutoff over momenta $p \gg 1/R$. In the denominator of Eq. (101), we can thus impose the large velocity condition $\alpha_p \ll pv$ in correspondence of $p \sim 1/R$, so that this condition is satisfied for all values of $p \leq 1/R$: this gives [with reference to Eq. (82)]

$$\tau_v \ll \tau_R. \tag{103}$$

Accordingly, the second threshold velocity $v_{II}$ is expected to be provided by the smallest velocity $v$ for which both conditions in Eq. (102) and (103) are satisfied, which depends on the particular choice of parameters and specific details of the system. In Eq. (101) we also recognize the term $\widetilde{V}_p \equiv V_p \exp[-Tp^2/(2\kappa)]$, featuring the Fourier transform of the interaction potential $V(x)$, corrected by a factor which takes into account the mean squared displacement of the particle in the harmonic trap induced by the thermal fluctuations [see also $\ell_T$ in Eq. (88)]. Accordingly, in this limit, the shadow becomes

$$\langle \varphi(x) \rangle_{\text{st}} \simeq -\frac{\lambda}{\gamma_\phi v} \int_{-\infty}^x \mathrm{d}x' \, \widetilde{V}(x'). \tag{104}$$

[Here we assumed that $\langle \varphi(x) \rangle_{\text{st}} \to 0$ for $|x| \to +\infty$, as expected from Fig. 3(a) in the main text.] As a consequence, the effective force term in Eq. (97) scales in this regime as $v^{-1}$, which produces an analogous dependence on $v$ for the average particle position $\langle Z \rangle$ in the steady state. Due to this very $v^{-1}$ dependence, we expect this *deterministic* effect to die out for very large $v$. In this last ("depinning") regime, the field is so slow with respect to the particle motion that no shadow can build up, and the particle effectively sees a rough landscape whose features are solely determined by the thermal fluctuations of the field. In order to describe this situation, we have to go back to the particle effective equation (91), the average over thermal fluctuations of which yields

$$\gamma_y \partial_t \langle \mathbf{Z} \rangle = -\kappa \langle \mathbf{Z} \rangle + \lambda \int \frac{\mathrm{d}^d p}{(2\pi)^d} i\mathbf{p} V_{-p} \left[ \left\langle \zeta_\mathbf{p}(t) e^{i\mathbf{p}\cdot\mathbf{Z}(t)} \right\rangle + \lambda V_p \int_0^\infty \mathrm{d}u \, \chi_\mathbf{p}(u) \left\langle e^{i\mathbf{p}\cdot[\mathbf{Z}(t)-\mathbf{Z}(t-u)]} \right\rangle \right]. \tag{105}$$

By using Novikov's theorem [35, 36]

$$\langle \zeta(t) F[\zeta] \rangle = \int \mathrm{d}s \, \langle \zeta(t) \zeta(s) \rangle \left\langle \frac{\delta F[\zeta]}{\delta \zeta(s)} \right\rangle, \tag{106}$$

where $F[\zeta]$ is any functional of the (Gaussian) noise $\zeta$, we can compute the expectation value

$$\left\langle \zeta_\mathbf{p}(t) e^{i\mathbf{p}\cdot\mathbf{Z}(t)} \right\rangle = i\mathbf{p} \int \mathrm{d}s \, C_\mathbf{p}(t-s) \left\langle e^{i\mathbf{p}\cdot\mathbf{Z}(t)} \frac{\delta \mathbf{Z}(t)}{\delta \zeta_{-p}(s)} \right\rangle$$

$$\simeq \frac{\lambda p^2 V_p}{\gamma_y} \int_{-\infty}^t \mathrm{d}s \, e^{-\kappa(t-s)/\gamma_y} C_\mathbf{p}(t-s) \left\langle e^{i\mathbf{p}\cdot[\mathbf{Z}(t)-\mathbf{Z}(s)]} \right\rangle_0, \tag{107}$$

where in the last step we used the equation of motion (91) of $\mathbf{Z}(t)$, we indicated by $\langle\ldots\rangle_0$ the expectation value over the unperturbed process with $\lambda = 0$, and we discarded higher-order terms in $\lambda$. The leading-order expression for the dynamical structure factor

$$\left\langle e^{i\mathbf{p}\cdot[\mathbf{Z}(t)-\mathbf{Z}(t-u)]} \right\rangle_0 \xrightarrow[t\to+\infty]{} e^{-p^2\sigma^2(u)} \tag{108}$$

was computed in Ref. [28], with $\sigma(u)$ given in Eq. (77). Setting $\partial_t \langle \mathbf{Z} \rangle = 0$ in Eq. (105) and taking the limit for $t \to +\infty$ on its r.h.s. then yields the steady-state average position

$$\langle \mathbf{Z} \rangle = \frac{\lambda^2}{\kappa} \int \frac{\mathrm{d}^d p}{(2\pi)^d} i\mathbf{p}|V_p|^2 \int_0^\infty \mathrm{d}u \left[ \chi_{\mathbf{p}}(u) + \frac{p^2}{\gamma_y} e^{-\kappa u/\gamma_y} C_{\mathbf{p}}(u) \right] e^{-p^2\sigma^2(u)} + \mathcal{O}(\lambda^4). \tag{109}$$

This last expression indeed coincides with Eq. (19) in the main text, *for any finite temperature $T > 0$*, which can be seen upon integrating by parts in $\mathrm{d}u$ and by using the fluctuation-dissipation theorem in the form of Eq. (76). It is simple to verify that the term proportional to $\chi_{\mathbf{p}}(u)$ in Eq. (109) decays as $1/v$ for large $v$, as it encodes the deterministic effect of the shadow. At large $v$, the only term that contributes is the one originating from the noise $\zeta_{\mathbf{p}}(t)$ and proportional to the field correlator $C_{\mathbf{p}}(u)$, which explicitly becomes

$$\langle \mathbf{Z} \rangle = \frac{\lambda^2 T}{\kappa \gamma_y} \int \frac{\mathrm{d}^d p}{(2\pi)^d} \frac{i\mathbf{p}\, p^2 |V_p|^2}{p^2 + r} \int_0^\infty \mathrm{d}u \exp\left[(i\mathbf{p}\cdot\mathbf{v} - \kappa/\gamma_y - \alpha_p)u - p^2\sigma^2(u)\right] + \mathcal{O}(\lambda^4). \tag{110}$$

This term is proportional to the strength $T$ of the thermal fluctuations and the coupling $\lambda^2$ to the field, in agreement with the interpretation we provided earlier in this section — i.e., that the features of the field landscape encountered by the particle in this regime are determined by the critical fluctuations of the former. The identification of the value $v_{III}$ of the velocity $v$ at which this regime begins is, however, not straightforward: it marks the value at which the field-induced thermal fluctuations start to dominate over its deterministic attraction. Since these fluctuations exhibit a rather strong dependence on the distance $r$ from the critical point [as it appears by inspecting the field correlator $C_{\mathbf{p}}(u)$ in Eq. (58)], then $v_{III}$ also shows a similar dependence on the correlation length $\xi = r^{-1/2}$ (as well as on the temperature $T$). In particular, Fig. 3(b) in the main text shows that $v_{III}$ decreases upon approaching the critical point $r = 0$, while the extents of the first and second velocity regimes decrease.

### C. The case of critical non-conserved dynamics

It is instructive, at this point, to inspect the case of critical non-conserved dynamics, i.e., to consider $r = 0$ and $\mathfrak{a} = 0$. In $d = 1$ and with a Gaussian interaction potential $V_p$ (with zero mean and variance $R$), the average particle position given in Eq. (19) of the main text can be simplified as

$$\langle Z \rangle = -\frac{\lambda^2 v}{\kappa} \int_0^\infty \frac{\mathrm{d}u}{\sqrt{4\pi}\Sigma(u)} \exp\left[-\frac{(vu)^2}{4\Sigma^2(u)}\right] \le 0, \tag{111}$$

$$\Sigma^2(u) \equiv R^2 + u/\gamma_\phi + \sigma^2(u). \tag{112}$$

The corresponding curve for the dissipation rate is plotted in Fig. 3(b) in the main text (solid green line with $\xi = \infty$). As the critical point is approached, we note that the crossover velocity $v_I$ described in Section VI B decreases; exactly at criticality, the entire low-velocity (adiabatic) regime disappears, in agreement with the expression of $v_I$ given in Eq. (100). Furthermore, for $r \to 0$, critical fluctuations are found to play a major role, as signaled by the fact that the crossover velocity $v_{III}$ also decreases considerably.

Interestingly, we note that the expression of $\langle Z \rangle$ for $r = 0$ in Eq. (112) has a finite, non-vanishing limit for $v \to 0$, while Eq. (19) of the main text, in the same limit but with a generic value of $r$, correctly vanishes independently of $r \ne 0$; in fact, at equilibrium, the presence of the field does not modify the equilibrium distribution of the particle [28]. We are thus led to the conclusion that the limits $r \to 0$ and $v \to 0$ do not commute: the physical interpretation is that, when $r \to 0$, the relaxation timescale $\tau_\xi$ of the field diverges [see Eq. (82)] and therefore the system is out of equilibrium for any finite value of $v$, no matter how small.

### VII. NUMERICAL SIMULATION OF THE STOCHASTIC DYNAMICS

In this Appendix we provide further details regarding the numerical simulation of the coupled stochastic dynamics of field and particle.

## A. Numerical integration scheme

The joint stochastic evolution of the field and the particle, given by Eqs. (2) and (3) in the main text, is integrated in time by using the stochastic Runge-Kutta method [37] with a fixed discretization time interval $\Delta t$. While the particle evolves in continuous space (i.e., off-lattice), the field is defined on $\Lambda = \left\{ \Delta x \sum_{i=1}^{d} \ell_i \mathbf{e}_i, 1 \leq \ell_i \leq L \right\}$, a cubic grid of $L^d$ lattice points spaced $\Delta x$ apart. We set out with $\phi(\mathbf{x}, t)$ and $\mathbf{Y}(t)$ and define the first auxiliary increment

$$\mathbf{K}_1(t) = -\Delta t \left[ \gamma_y^{-1} \sum_{\mathbf{x} \in \Lambda} V(\mathbf{x} - \mathbf{Y}(t)) \nabla \phi(\mathbf{x}, t) + \gamma_y^{-1} \kappa (\mathbf{Y} - \mathbf{v}t) \right] + \sqrt{2\gamma_y^{-1} T \Delta t} \, \boldsymbol{\nu}(t) \,. \tag{113}$$

Here, the field gradient is approximated by its lattice discretization,

$$(\nabla \phi(\mathbf{x}))_i \to [\phi(\mathbf{x} + \Delta x \, \mathbf{e}_i) - \phi(\mathbf{x} - \Delta x \, \mathbf{e}_i)] / (2\Delta x) \,. \tag{114}$$

Further, $\boldsymbol{\nu}(t)$ denotes an uncorrelated Gaussian random vector with unit variance. The first increment $\mathbf{K}_1$ is added to $\mathbf{Y}(t)$ to obtain the auxiliary value $\hat{\mathbf{Y}}(t) = \mathbf{Y}(t) + \mathbf{K}_1(t)$. The stochastic increment of the field is then evaluated by using the auxiliary value as

$$\phi(\mathbf{x}, t + \Delta t) = -\Delta t \left[ \gamma_\phi^{-1} (\nabla^2 + r) \phi(\mathbf{x}, t) - V(\mathbf{x} - \hat{\mathbf{Y}}(t)) \right] + \sqrt{2\gamma_\phi^{-1} T \Delta t} \, \boldsymbol{\eta}(\mathbf{x}, t), \tag{115}$$

where the Laplacian is approximated via

$$\nabla^2 \phi(\mathbf{x}, t) = \sum_{i=1}^{d} [\phi(\mathbf{x} + \Delta x \, \mathbf{e}_i, t) - 2\phi(\mathbf{x}, t) + \phi(\mathbf{x} - \Delta x \, \mathbf{e}_i, t)] / (2\Delta x)^2. \tag{116}$$

Hereafter, the position of the particle is updated using the new field configuration by defining the second auxiliary increment

$$\mathbf{K}_2(t) = -\Delta t \left[ \gamma_y^{-1} \sum_{\mathbf{x} \in \Lambda} V(\mathbf{x} - \mathbf{Y}(t)) \nabla \phi(\mathbf{x}, t + \Delta t) + \gamma_y^{-1} \kappa (\mathbf{Y} - \mathbf{v}t) \right] + \sqrt{2\gamma_y^{-1} T \Delta t} \, \boldsymbol{\nu}(t). \tag{117}$$

The noise $\boldsymbol{\nu}(t)$ used in this second increment is identical to the one previously used in Eq. (113). Finally, $\mathbf{Y}(\mathbf{t})$ is updated as

$$\mathbf{Y}(t + \Delta t) = \mathbf{Y}(t) + \frac{1}{2} [\mathbf{K}_1(t) + \mathbf{K}_2(t)]. \tag{118}$$

This scheme can be suitably adapted in order to to include, e.g., conserved field dynamics. As a second-order stochastic Runge-Kutta integration scheme, the local truncation error in time is of order $\mathcal{O}(\Delta t^3)$, and can be improved by using higher-order schemes [37].

The initial condition $\phi(\mathbf{x}, t=0) \equiv \phi(\mathbf{x})$ of the field $\phi(\mathbf{x}, t)$ is drawn from the Gaussian equilibrium distribution of the decoupled field (i.e., with $\lambda = 0$, see Section IV B). By applying a $d$-dimensional Fourier transform to $\phi$, one obtains

$$\phi(\mathbf{x}) = \frac{1}{(2\pi)^d} \sum_{\mathbf{p} \in \Lambda^{-1}} e^{i\mathbf{p}\mathbf{x}} \phi_\mathbf{p}, \tag{119}$$

where $\mathbf{p}$ sums over the reciprocal lattice $\Lambda^{-1} = \left\{ \frac{2\pi}{\Delta x} \sum_{i=1}^{d} p_i \mathbf{e}_i, -L/2 + 1 \leq p_i \leq L/2 \right\}$. We randomly draw the complex modes $\phi_\mathbf{p} \in \mathbb{C}$ for momenta belonging to the positive half-cube $\mathbf{p} \in \left\{ \mathbf{p}' \in \Lambda^{-1} | p'_1 \geq 0 \right\}$, containing all vectors in the reciprocal lattice with positive first component; while the magnitude is drawn from a Gaussian random distribution $\mathcal{N}$, i.e., $\|\phi_\mathbf{p}\| \sim \mathcal{N}\left(0, \frac{2\pi}{(L\Delta x)^2} (p^2 + r)\right)$, the complex phase $\theta \in \mathcal{U}([0, 2\pi))$ is drawn from a uniform random distribution $\mathcal{U}$, leading to a random initial mode $\phi_\mathbf{p} = \|\phi_\mathbf{p}\| e^{i\theta}$. If $r = 0$, we set $\phi_{\mathbf{p}=0} = 0$. Then, to ensure that $\phi(\mathbf{x})$ is real, we set $\phi_{-\mathbf{p}} = \phi_\mathbf{p}^*$ for the remaining vectors in the negative half-cube. The initial condition then follows from inserting the random modes into Eq. (119). The particle is initialised at the centre of the trap at $t = 0$. Since neither the particle nor the field start in their joint equilibrium distribution ($\lambda \neq 0$), we allow for thermalization by evolving the system for typically $10^4$ timesteps at a time discretization of $\Delta t = 10^{-2}$, after which we assume the system to have settled into a (nonequilibirium) steady state.

## B. Choice of parameters

Here we discuss the choice of the values of the parameters of the model to be used in the numerical simulations, so that they eventually correspond to the typical time and length scales observed in experiments with colloidal particles. As a prototypical example, we consider the case of a $\mu$m–sized colloidal particle immersed in a binary liquid mixture close to its demixing transition [38–42], so that the field $\phi$ represents the relative concentration of the two species that compose the mixture. It must be emphasized, however, that our model is not meant to reproduce quantitatively the results of such experiments (for which hydrodynamic effects would need to be taken into account), but rather to highlight the role played by spatial correlations. Moreover, other very diverse physical systems (such as inclusions in lipid membranes [43–46], microemulsions [47–50], or defects in ferromagnetic systems [51–56]) also fall within the scopes of our model, and they may entail very different time and length scales.

Binary liquid mixtures can undergo a demixing phase transition close to room temperatures [57], in correspondence of which the correlation length $\xi = \xi(T)$ diverges. In real experiments, achieving large values of $\xi$ is generally challenging, as it requires a fine-tuning and highly accurate control of the temperature throughout the sample. However, in the following we will assume that $\xi$ can indeed be made much larger than the particle radius $R$, in order to emphasize the qualitative effects of spatially extended correlations, which are the central topic of this paper.

Note that our model features several parameters, but only a limited number of physical units (i.e., mass, length, time and temperature). As detailed in Ref. [30], close to the critical point, the system can be conveniently described in terms of a reduced set of dimensionless parameters, which correspond to ratios of the typical time and length scales of the system described in Section VI A. These parameters are

$$w \equiv \frac{\tau_R}{2\tau_v}, \qquad \rho \equiv \tau_R/\tau_\kappa, \qquad g \equiv \frac{\lambda^2}{\kappa R^d}, \qquad \text{and} \qquad \epsilon \equiv \frac{l}{R}. \tag{120}$$

The first is the Weissenberg number $w = \text{Wi}$, which compares the shear rate due to dragging with the typical relaxation time of the field $\tau_R$, over length scales of the order of the particle radius $R$. The second parameter, $\rho$, compares the relaxation time of the field with the time scale $\tau_\kappa$ set by the harmonic trap. The effective coupling $g$ quantifies the strength of the interaction between the particle and the field. Finally, $\epsilon$ is the ratio between the *thermal length* $l = \sqrt{k_B T/(2\kappa)}$ (i.e., the mean-squared displacement of the particle in the trap according to equipartition in equilibrium) and the particle radius $R$. Specializing these expressions to the case of critical non-conserved dynamics in $d = 2$ (see Section VI A and Ref. [30]) yields $g = \lambda^2/(\kappa R^2)$, $w = Rv\gamma_\phi/2$, and $\rho = \kappa R^2 \gamma_\phi/\gamma_y$.

Having identified the relevant parameters in Eq. (120), we now discuss which values of them are within experimental reach. The typical magnitude of the thermal length $l$ can be estimated at room temperature $T \simeq 300\,\text{K}$ by assuming a typical value of the optical trap strength $\kappa \simeq 0.5\,\text{pN}/\mu\text{m}$ [40], yielding $l \simeq 100\,\text{nm}$, and hence $\epsilon \simeq 0.1$ for a $\mu$m-sized particle. Next, the typical relaxation time scale $\tau_\phi$ of a near-critical binary mixture can be estimated by using model H [58]. Within mode-coupling theory and for momenta $p$ such that $p\xi \ll 1$ (i.e., for small $\xi$, which has been the case in past experiments [38–42]), the relaxation timescale of the field is given by [59]

$$\tau_\phi^{-1}(p) \simeq D_\xi \, p^2, \tag{121}$$

where

$$D_\xi = \frac{k_B T}{6\pi\eta\xi}, \tag{122}$$

and $\eta$ is the fluid viscosity. By comparison with the diffusion coefficient $D_R$ of the particle [42]

$$D_R = \frac{k_B T}{6\pi\eta R} \simeq 0.22(\mu\text{m})^2\text{s}^{-1}, \tag{123}$$

one can estimate the relaxation time scale of the field over distances of the order of the particle radius as

$$\tau_R \simeq \tau_\phi(p \sim \pi/R) \simeq \frac{R\xi}{\pi^2 D_R} \simeq 10\,\text{ms}, \tag{124}$$

where we chose $\xi \simeq R/50$ (which is common for water-lutidine mixtures [41, 42]). Since the typical relaxation time $\tau_\kappa$ of optical traps is of the order of a few tens of ms [41], this gives $\rho \simeq 10^{-1}$. This value is expected to increase if the correlation length $\xi$ can be made larger, since this generally entails longer relaxation times $\tau_\phi$ [20]. Next, with typical drag speeds up to a few tens of $\mu$m/s [60], one can explore small Weissenberg numbers up to $w \simeq 10^{-1}$. On the other hand, the effective coupling $g$ quantifies the strength of the interaction between the particle and the field,

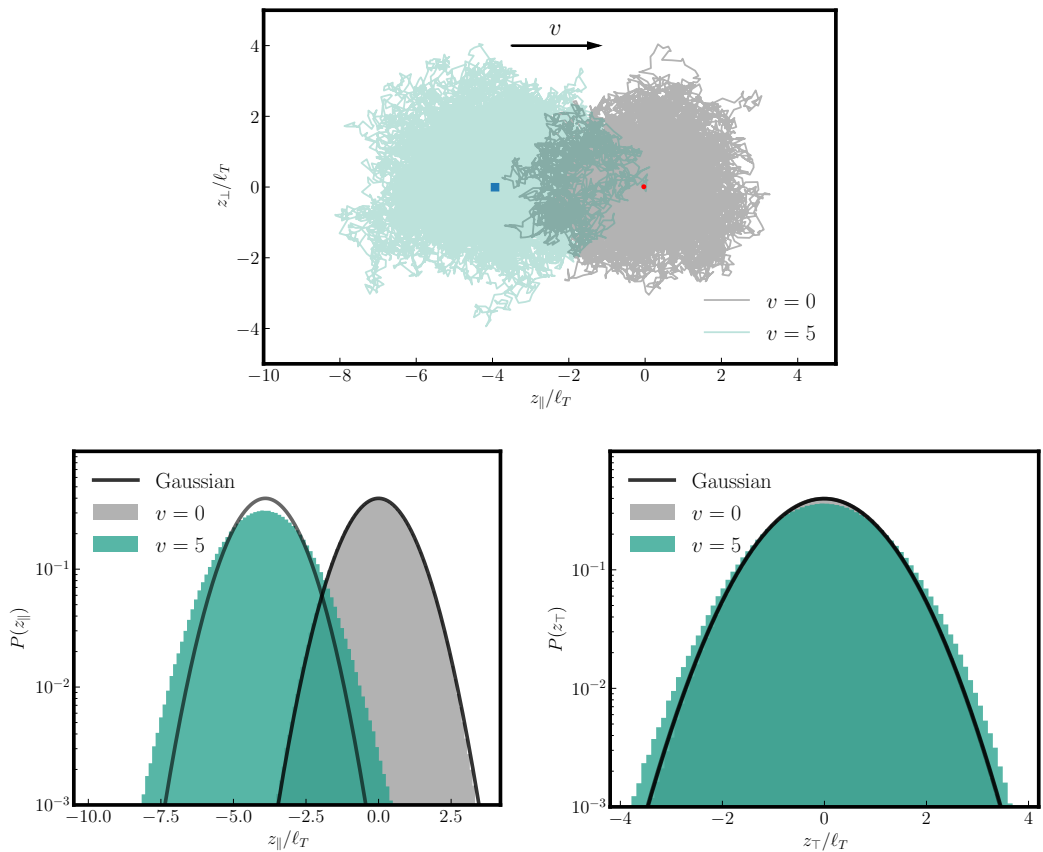

FIG. 1. **Upper panel:** scatter plot of the particle position, in the comoving frame, as measured in numerical simulations in $d = 2$ and for a field with non-conserved dynamics. The average positions for $v = 0$ and $v = 5$ are indicated with a red dot and a blue square, respectively. **Lower panels:** histograms of the particle position along the directions parallel ($z_\parallel$) and perpendicular ($z_\perp$) to the trap velocity **v**. In the simulations we used the parameter set given in Eq. (125), with $\ell_T$ given in Eq. (88). In the plot, we indicated with a solid black line the Gaussian distribution, which is recovered for $v = 0$ (boundary of the gray regions) or in the absence of particle-field interactions, i.e., for $\lambda = 0$ (leftmost curve in the left panel).

and its amplitude thus depends on the specific coupling mechanism realized in a certain experiment. At present, it is still unclear how to estimate its order of magnitude on the basis of past experimental data, but we clearly expect any overall qualitative effect to be enhanced if $g$ can be made larger. We finally keep an eye for consistency on the ratio $\Theta \equiv \tau_v/\tau_\kappa = \rho/(2w)$ [see Eq. (120)] of the shear rate to the trap time scale which, based on the estimates above for $\rho$ and $w$, is typically of $\mathcal{O}(1)$.

The experimentally accessible values of the parameters in Eq. (120) discussed above are reproduced in numerical simulations by choosing the following set of values of the model parameters:

$$T = 0.7, \quad \kappa = 0.2, \quad v = 5, \quad R = 2, \quad \lambda = 10, \quad r = 10^{-4}, \quad \gamma_\phi^{-1} = 20, \quad \text{and} \quad \gamma_y^{-1} = 5, \qquad (125)$$

where space is expressed in units of the lattice spacing $\Delta x$, and we furthermore chose a discretized time step $\Delta t = 10^{-2}$. This choice corresponds, in fact, to $w = 0.25$, $\rho = 0.2$, $\epsilon = 0.66$, $\Theta = 0.4$, and $g = 125$. The lattice size $L = 128$ is chosen sufficiently large so as to accommodate the tails of the field shadow [see Fig. 3(a) in the main text], and to avoid spurious field currents due to the *stirring* that the particle would generate if it were dragged around the periodic boundaries [30].

### C. Numerical measurement of the distribution of the particle position

Using the numerical simulation described above, we determined the statistics of the particle position, shown in Fig. 1. In particular, for a field in $d = 2$ with non-conserved dynamics, the upper panel of Fig 1 presents a scatter

plot of the particle position **Z** measured in the comoving frame. This shows clearly two of the main features already emerging from the cumulant generating function, predicted perturbatively (i.e., for small $\lambda$) in Eq. (18) of the main text, namely:

1. At finite $v$, the steady-state average position (blue square in Fig. 1) lags behind the one obtained for $v = 0$ (red dot). Note that the latter coincides with the average position found at $\lambda = 0$, i.e., $\langle \mathbf{Z} \rangle = 0$, because the coupling to the field does not modify the one-time properties of the particle position statistics at equilibrium (this result was proven non-perturbatively in Ref. [28]).

2. The variance of the particle position increases in the directions parallel and perpendicular to **v** by a different amount. Note that in the absence of the field (i.e., for $\lambda = 0$) the particle variance would be $\langle Z_l Z_m \rangle_c = (T/\kappa)\delta_{lm}$ [see Eq. (81)], independently of the value of $v$. Accordingly, the observed anisotropic change of the variance of the particle position is the result of the combined effect of the presence of the field and of the dragging.

The lower panels of Fig. 1 show histograms of the distribution of the particle positions $z_\parallel$ and $z_\perp$ along the directions parallel (left) and perpendicular (right) to the trap velocity **v**, respectively, compared to the case with $\mathbf{v} = 0$ (gray shading). Note that the distribution for $\mathbf{v} \neq \mathbf{0}$ is non-Gaussian, although the non-Gaussian character is not immediately apparent for the values of the parameters used in this figure. However, this non-Gaussianity is evident at the lowest perturbative order in $\lambda$, as we discuss in the main text [see Eq. (18) of the main text].

### D. Numerical measurement of the heat dissipation field

In Eq. (9) of the main text, we introduced the field $đQ_\phi$ of heat dissipation and we determined it numerically for a Gaussian field $\phi$ evolving in $d = 2$ according to a non-conserved dynamics (i.e., $\mathfrak{a} = 0$). This heat dissipation field was shown in Fig. 2 of the main text. Here we provide supplementary information on the numerical aspects of its determination.

In order to measure the heat field, we evaluate Eq. (9) of the main text for non-conserved Gaussian dynamics. For the heat $đQ_\phi$ dissipated within a time interval $[t, t + \Delta t]$ we find

$$đQ_\phi = \left[ \frac{\delta \mathcal{H}_\phi}{\delta \phi} + \frac{\delta \mathcal{H}^{\text{int}}}{\delta \phi} \right] \circ d\phi(\mathbf{x}, t)$$
$$= \left[ (\nabla^2 + r)\phi(\mathbf{x}, t) + \lambda V(\mathbf{x} - \mathbf{Y}(t)) \right] \circ d\phi(\mathbf{x}, t), \tag{126}$$

with the increment $d\phi(\mathbf{x}, t)$ numerically approximated by the forward difference $\phi(\mathbf{x}, t + \Delta t) - \phi(\mathbf{x}, t)$. We are interested in the *average* heat dissipated in the nonequilibirum steady state, i.e., in $\langle đQ_\phi \rangle$. As discussed in Sec. I B, the average $\langle (\nabla^2 + r)\phi(\mathbf{x}, t) \circ d\phi \rangle$ of the first contribution on the r.h.s. of Eq. (126) vanishes in the long-time limit. Accordingly, it is therefore sufficient to evaluate the average of the second term. It remains to evaluate this second term in the comoving frame in which the trap is at rest, and where the particle and the field are described by **Z** and $\varphi$ [see Eqs. (16) and (17) in the main text], respectively. Applying the Galilei transformations introduced in the main text, the average heat in the comoving frame is then evaluated as

$$\langle đQ_\varphi \rangle (\mathbf{z}) = (\lambda/2) \langle [V(\mathbf{z} - \mathbf{Z}(t)) + V(\mathbf{z} - \mathbf{Z}(t + \Delta t))] \cdot [\varphi(\mathbf{z}, t + \Delta t) - \varphi(\mathbf{z}, t)] \rangle . \tag{127}$$

In order to sample the nonequilibrium steady state, we average both in time and over an ensemble. Following the procedure described in Sec. VII A, we generate $M \simeq 10^3$ trajectories of $\mathbf{Z}^{(m)}$ and $\varphi^{(m)}$ ($m = 1, ..., M$) of typically $N \simeq 10^5$ to $10^6$ timesteps with a discretization of $\Delta t = 10^{-2}$. The initial condition of the field is chosen as prescribed in Sec. VII A, while the particle is initialized at $\mathbf{Z}(t = 0) = 0$. We discard the first $T_{\text{th}} = 10^4$ timesteps of the numerical integration to allow the system to reach the nonequilibrium steady state in the comoving frame. This duration is determined by inspection of $\langle Z \rangle$, ensuring it has converged to its steady state. The average over both time and ensemble then reads

$$\langle đQ_\varphi(\mathbf{z}) \rangle \tag{128}$$
$$= \frac{1}{M} \sum_{m=1}^{M} \left\{ \frac{1}{N - T_{\text{th}}} \sum_{j=T_{\text{th}}}^{N} \lambda \frac{V\left(\mathbf{z} - \mathbf{Z}^{(m)}(j\Delta t)\right) + V\left(\mathbf{z} - \mathbf{Z}^{(m)}((j+1)\Delta t)\right)}{2} \left[ \varphi^{(m)}(\mathbf{z}, (j+1)\Delta t) - \varphi^{(m)}(\mathbf{z}, \Delta t) \right] \right\}.$$

Dividing the result by $\Delta t$, one obtains the numerically estimated average dissipation rate $\langle \dot{Q}_\varphi(\mathbf{z}) \rangle$ shown in Fig. 2 in the main text.

---




\end{document}